\documentclass[preprint,authoryear,12pt,3p,a4paper]{elsarticle1}

\usepackage[latin1]{inputenc}

\usepackage{graphicx}

\newcommand{\incfig}[2] {\includegraphics[width=#1]{#2}}
\newcommand{\rmd}{{\rm d}}
\newcommand{\rme}{{\rm e}}
\newcommand{\gd}{{\ast}}
\newcommand{\outline}[1]{}

\newcommand{\D}{\mathrm{D}}
\newcommand{\m}{\mathrm{m}}

\newcommand{\PD}{P_\mathrm{D}}
\newcommand{\PDk}{P_{\mathrm{D},k}}

\journal{$ $}

\begin{document}

\title{Dependence of defaults and recoveries in structural\\ credit risk models}

\author{Rudi Schäfer}
\ead{rudi.schaefer@uni-duisburg-essen.de}
\address{Faculty of Physics, University of Duisburg-Essen, Germany}

\author{Alexander F. R. Koivusalo}
\address{Danske Capital, Copenhagen, Denmark}


\begin{abstract}
The current research on credit risk is primarily focused on modeling default probabilities. 
Recovery rates are often treated as an afterthought; they are modeled independently, in many cases they are even assumed constant. This is despite of their pronounced effect on the tail of the loss distribution.
Here, we take a step back, historically, and start again from the Merton model, where defaults and recoveries are both determined by an underlying process. Hence, they are intrinsically connected.
For the diffusion process, we can derive the functional relation between expected recovery rate and default probability. This relation depends on a single parameter only.
In Monte Carlo simulations we find that the same functional dependence also holds for jump--diffusion and GARCH processes.
We discuss how to incorporate this {\it structural recovery rate} into reduced--form models, in order to restore essential structural information which is usually neglected in the reduced--form approach.
\vspace*{2ex}
\end{abstract}


\maketitle

\section*{Introduction}  \label{sec1}

An accurate description of portfolio credit risk is of vital interest for any financial institution.
It is also a prerequisite for realistic ratings of structured credit derivative products.
Furthermore, it is a crucial aspect in banking regulations.
We can distinguish two conceptually different approaches to credit risk modelling: structural and reduced--form approaches.
The structural models go back to \cite{black73} and \cite{Merton74}.
The Merton model assumes that a company has a certain amount of zero--coupon debt which becomes due at a fixed maturity date. The market value of the company is modelled by a stochastic process. A possible default and the associated recovery rate are determined directly from this market value at maturity.
In the reduced--form approach default probabilities and recovery rates are described independently by stochastic models. The aim is to describe the dependence of these quantities on common (macroeconomic) covariates or risk factors. 
For some well known reduced--form model approaches see, e.g., \cite{JarrowTurnbull95}, \cite{JarrowLandoTurnbull97}, 
\cite{DuffieSingleton99}, \cite{HullWhite2000} and  \cite{Schoenbucher2003}.
First Passage Models constitute a third approach --- which is usually regarded as structural, but is better described as a mixed or pseudo--structural approach. They were first introduced by \cite{BlackCox1976}.
As in the Merton model, the market value of a company is modelled as a stochastic process.
Default occurs as soon as the market value falls below a certain threshold.
In contrast to the Merton model, default can occur at any time.
In this approach, the recovery rate is not determined by the underlying process for the market value.
Instead recovery rates are modelled independently, for example, by a reduced--form approach (see e.g., \cite{AsvanuntStaal09a, AsvanuntStaal09b}).  
In some cases recovery rates are even assumed constant, for instance, in \cite{Giesecke2004}. 
The independent modelling of default and recovery rates leads to a serious underestimation of large losses. 

In this paper we take a step back, historically, and revisit the Merton model, where defaults and recoveries are both determined by an underlying process. Hence, they are intrinsically connected.
For a correlated diffusion process the Merton model has been treated analytically, e.g., in \cite{bluhm02} and in \cite{Giesecke2004}.
In a straightforward calculation we can also derive the functional relation between expected recovery rate and default probability. This relation depends on a single parameter only.
In Monte Carlo simulations we find that the same functional dependence also holds for other processes like jump--diffusion and GARCH.
We discuss how to incorporate this relation into reduced--form models, in order to restore essential structural information which is usually neglected in the reduced--form approach.

The paper is organized as follows. We give a short introduction to the Merton model in Section~\ref{sec2}. In Section~\ref{sec3} we treat the diffusion case analytically and compare the results to Monte--Carlo simulations. In Sections~\ref{sec4} and \ref{sec5} we extend the Merton model to jump--diffusion and GARCH processes. We discuss the applicability of the structural recovery rate beyond the Merton model in Section~\ref{sec7}. We summarize our findings in Section~\ref{sec7}.

\section{Merton model} \label{sec2}

The Merton model assumes that a company $k$ has a certain amount of zero--coupon debt;
this debt has the face value $F_k$ and will become due at maturity time $T$. 
The company defaults if the value of its assets at time $T$ is less than the face value, i.e., if $V_k(T)<F_k$.
The recovery rate then reads $R_k=V_k(T)/F_k$ and the loss given default is 
\begin{equation}\label{lossgd}
L_k^\gd=1-R_k=\frac{F_k-V_k(T)}{F_k}  \;.
\end{equation}
We denote the loss given default with an asterisk to distinguish it from the loss including non--default events.
The individual loss can be expressed as 
\begin{equation}\label{eq:lossk}
L_k=\left( 1-\frac{V_k(T)}{F_k} \right) \, \Theta\left( 1-\frac{V_k(T)}{F_k} \right) \;,
\end{equation}
where $\Theta$ is the Heaviside function.
In the Merton model, defaults and losses --- and hence also recoveries --- are directly determined by the asset value at maturity. 
Therefore, the stochastic modelling of the market value $V_k(t)$ of a company allows to assess its credit risk.
Let $p_{V_k}(V_k(T))$ be the probability density function (pdf) of the market value at maturity. 
Then the default probability is given by
\begin{equation}
\PDk=\int\limits_0^{F_k} p_{V_k}(V_k(T))\, \rmd V_k(T)
\end{equation}
and the expected recovery rate can be calculated as
\begin{equation}
\left< R_k \right> = \frac{1}{\PDk} \int\limits_0^{F_k} \frac{V_k(T)}{F_k}\ p_{V_k}(V_k(T))\, \rmd V_k(T) \;.
\end{equation}

Let us now consider a portfolio of $K$ credit contracts, where the market value of each company $k$ is correlated to one or more covariates.
Conditioned on the values of the covariates we obtain different values for $\PDk$ and $\left< R_k \right>$. In fact, we find a functional dependence between default probability and recovery rate. This is in stark contrast to modelling approaches which assume an independence of these quantities.
In section \ref{sec3} we derive this functional dependence analytically for the diffusion case.
We compare the result with Monte Carlo simulations. In section \ref{sec4} we extend the model by including jump terms in the random process. In section \ref{sec5} we consider a GARCH process. In both cases we find the same functional dependence between default and recovery rates.

\section{Correlated diffusion} \label{sec3}

\subsection{Analytical discussion}
In the diffusion case we can easily derive all results analytically. 
To keep the notation simple we consider a homogeneous portfolio of size $K$ with the same parameters for each asset process, and with the same face value, $F_k=F$, and initial market value, $V_k(0)  =V_0$.
We model the time evolution of the market value of a single company $k$ 
by a stochastic differential equation of the form
\begin{equation} \label{eqDiff}
\frac{\rmd V_k}{V_k} = \mu \rmd t +  \sqrt{c}\, \sigma \, \rmd W_{\rm m} 
 + \sqrt{1-c}\, \sigma \, \rmd W_k \ .
\end{equation}
This is a correlated diffusion process with a deterministic term $\mu\rmd t$ and
a linearly correlated diffusion. The parameters of this process are the drift constant $\mu$, the volatility $\sigma$ and the correlation coefficient $c$. 
The Wiener processes $\rmd W_k$ and 
$\rmd W_{\rm m}$ describe the idiosyncratic and the market fluctuations, respectively.


For discrete time increments $\Delta t=T/N$, where the time to maturity $T$ is divided into $N$ steps, we arrive at the discrete formulation of Eq.~(\ref{eqDiff}). The market value of company $k$ at maturity can be written as
\begin{equation}\label{eq:VkTdiff}
V_k(T)=V_0 \prod\limits_{t=1}^{N} 
\left(1+ \mu \Delta t +  \sqrt{c}\, \sigma \eta_{{\rm m},t} \sqrt{\Delta t} 
 + \sqrt{1-c}\, \sigma \varepsilon_{k,t} \sqrt{\Delta t} \right) \;.
\end{equation}
We define the market return $X_{\rm m}$ as the average return of all single companies $k$ over the time horizon up to maturity,
\begin{eqnarray}
X_{\rm m} & = & \frac{1}{K}\sum\limits_{k=1}^K \left( \frac{V_k(T)}{V_0}-1 \right) \\
 & = & \frac{1}{K}\sum\limits_{k=1}^K \prod\limits_{t=1}^{N} 
\left(1+ \mu \Delta t +  \sqrt{c}\, \sigma \eta_{{\rm m},t} \sqrt{\Delta t} 
 + \sqrt{1-c}\, \sigma \varepsilon_{k,t} \sqrt{\Delta t} \right) -1 \;.
\end{eqnarray}
For $K\to\infty$ we can express the average over $k$ as the expectation value for $\varepsilon_{k,t}$. Due to the independence of $\varepsilon_{k,t}$ for different $k$ and $t$, we can write
\begin{equation}\label{eqXmdiff}
X_{\rm m} + 1 =  \prod\limits_{t=1}^{N} 
\left(1+ \mu \Delta t +  \sqrt{c}\, \sigma \eta_{{\rm m},t} \sqrt{\Delta t} 
 + \sqrt{1-c}\, \sigma \langle \varepsilon_{k,t} \rangle \sqrt{\Delta t} \right) \;,
\end{equation}
with
\begin{equation}
\langle \varepsilon_{k,t} \rangle = 0   \;.
\end{equation}

\begin{figure*}
\begin{center}
 \incfig{0.49\textwidth}{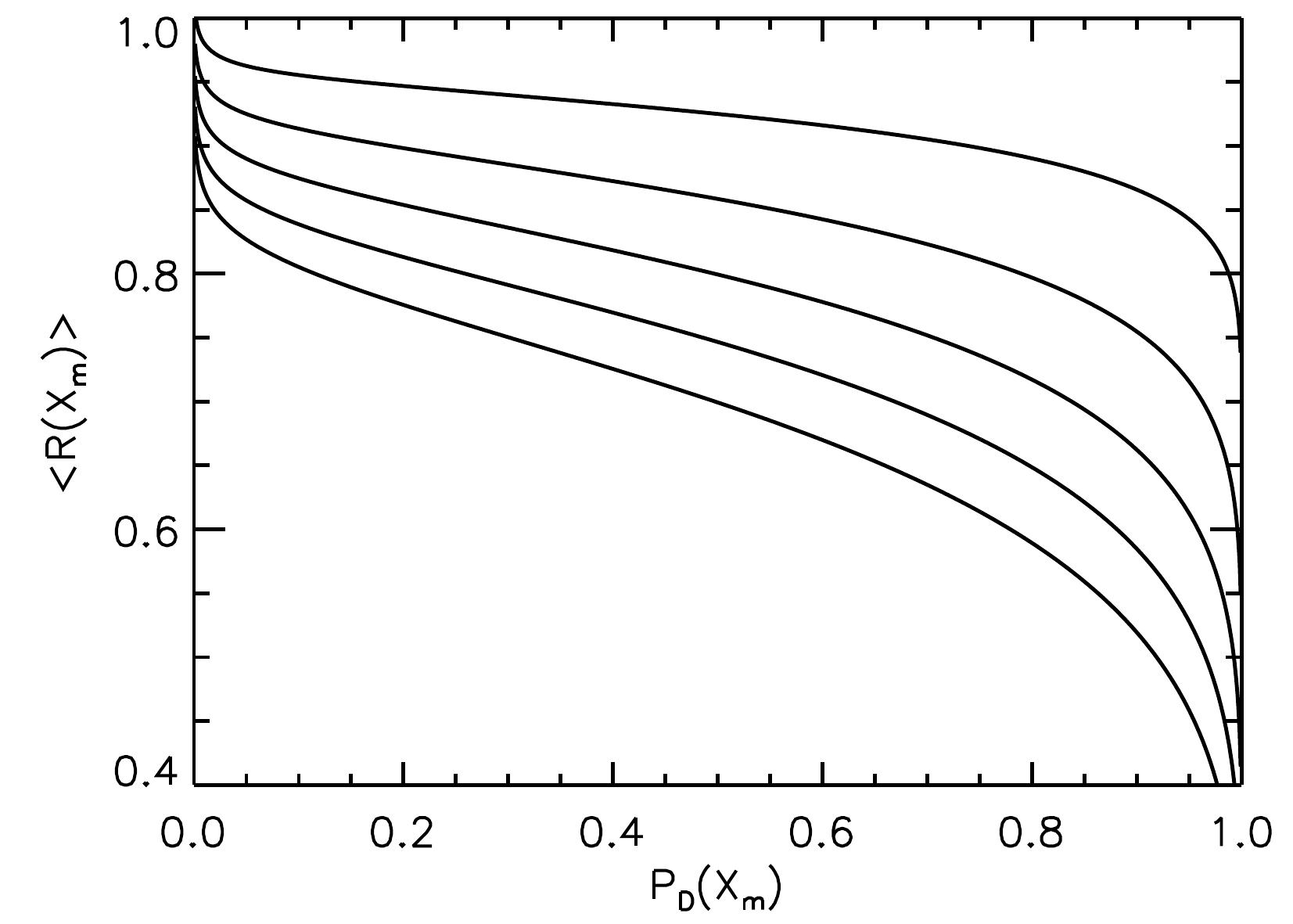}
 \incfig{0.49\textwidth}{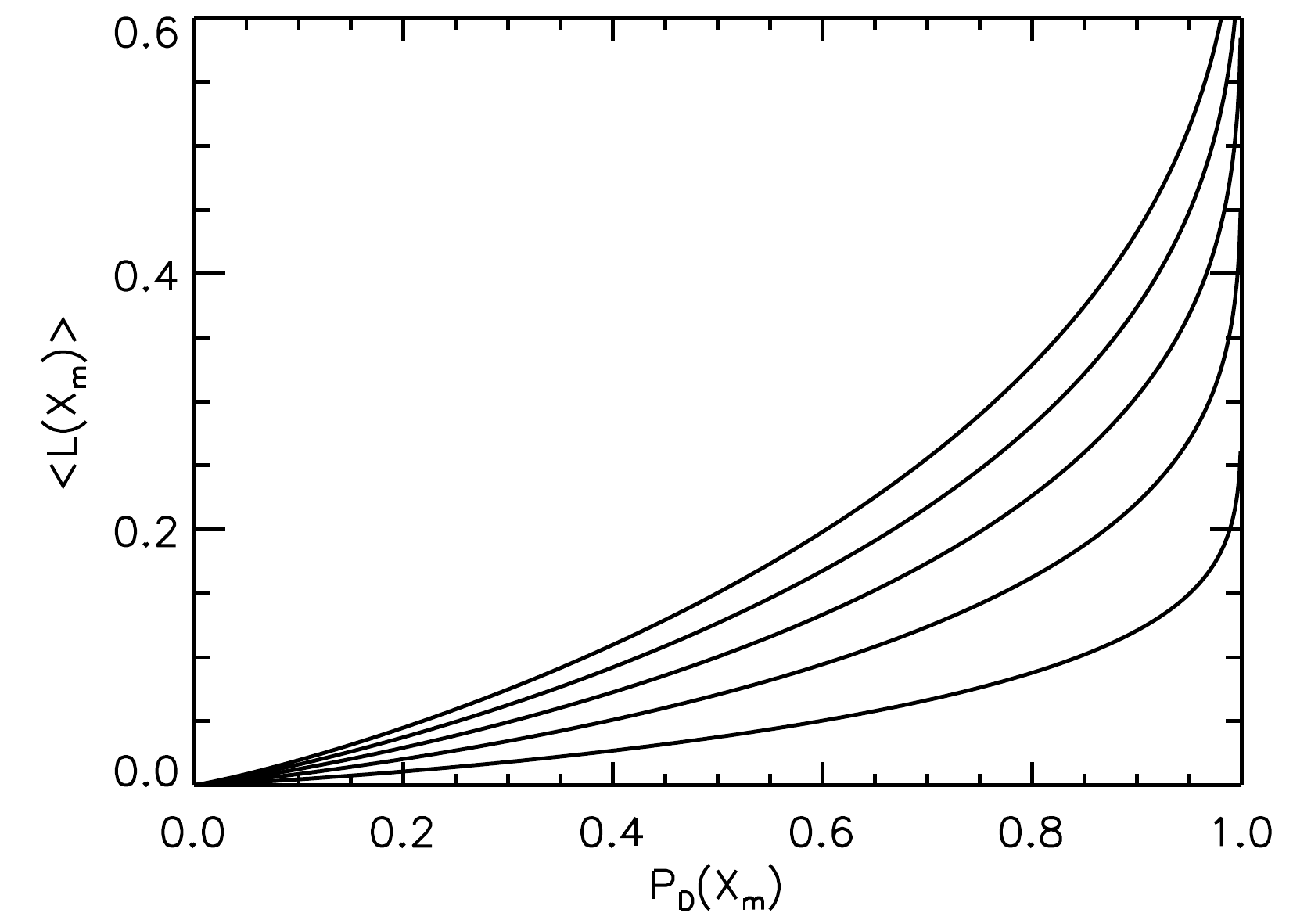}
\caption{
Parameter dependence of the functional relations $\langle R(\PD) \rangle$ (left plot) 
and $\langle L(\PD) \rangle$ (right plot).
The parameter values are $B=0.1$, 0.2, 0.3, 0.4 and 0.5 (top to bottom in left plot, and bottom  to top  in right plot).
}
\label{fig:b-dependence}
\end{center}
\end{figure*}

Thus, expression~(\ref{eqXmdiff}) simplifies to
\begin{eqnarray} \label{eq:Xmapprox}
X_{\rm m} + 1 &=& \prod\limits_{t=1}^{N} \left(1+ \mu \Delta t +  \sqrt{c}\, \sigma \eta_{{\rm m},t} \sqrt{\Delta t}\, \right)   \nonumber \\
&=& \exp\left(\sum\limits_{t=1}^{N} \ln\left(1+ \mu \Delta t +  \sqrt{c}\, \sigma \eta_{{\rm m},t} \sqrt{\Delta t}\, \right) \right)  \nonumber \\
&\approx& \exp\left( \left(\mu-\frac{c\,\sigma^2}{2}\right) T +  \sigma\sqrt{c \Delta t} \sum\limits_{t=1}^{N} \eta_{{\rm m},t}  \right) \;.
\end{eqnarray}
In the last step of the calculation we expanded the logarithm up to first order in $\Delta t$. 
The random variables $\eta_{{\rm m},t}$ are standard normal distributed. Therefore the variable
\begin{equation}
\ln\left( X_{\rm m} + 1 \right) =
\left(\mu-\frac{c\,\sigma^2}{2}\right) T +  \sigma\sqrt{c T} \frac{1}{\sqrt{N}} \sum\limits_{t=1}^{N} \eta_{{\rm m},t}  
\end{equation}
is normal distributed with mean $\mu T-\frac{1}{2}c\sigma^2T$ and variance $c\sigma^2T$.
This implies a shifted log-normal distribution for the market return itself,
\begin{equation}\label{eq:pofXm}
p_{X_{\rm m}}(X_{\rm m})=\frac{1}{ (X_{\rm m}+1) \sqrt{2\pi c\sigma^2 T}} 
\exp\left( -\frac{\left(\ln(X_{\rm m}+1) - \mu T + \frac{1}{2}c\sigma^2 T  \right)^2}{2c\sigma^2 T}   \right)   \;.
\end{equation}
For a single company $k$ we can write
\begin{eqnarray}
\ln\frac{V_k(T)}{V_0} &=& \sum\limits_{t=1}^N \ln\left( 1+ \mu \Delta t +  \sqrt{c}\, \sigma \eta_{{\rm m},t} \sqrt{\Delta t} + \sqrt{1-c}\, \sigma \varepsilon_{k,t} \sqrt{\Delta t} \right) \\
&\approx&  \ln\left( X_{\rm m} + 1 \right) - \frac{(1-c)\sigma^2}{2}T + \sigma\sqrt{(1-c) T} \frac{1}{\sqrt{N}} \sum\limits_{t=1}^{N} \varepsilon_{k,t} \;.
\end{eqnarray}
Conditioned on a fixed value for the market return $X_{\rm m}$, all variables $V_k(T)$ are independent and $\ln V_k(T)/V_0$ is normal distributed with mean $\ln\left( X_{\rm m} + 1 \right) - \frac{1}{2}(1-c)\sigma^2T$ and variance $(1-c)\sigma^2T$.
Since we consider a homogeneous portfolio, we will omit the index $k$ in the following. This allows for a clearer notation.
The probability density function for the market value $V(T)$ is given by the following log-normal distribution
\begin{equation}\label{eq:pdfVk}
p_{V}(V(T))= \frac{1}{V(T) \sqrt{2\pi (1-c)\sigma^2 T}} 
\exp\left( -\frac{\left(\ln\frac{V(T)}{V_0} -  \ln\left( X_{\rm m} + 1 \right) + \frac{1}{2}(1-c)\sigma^2 T  \right)^2}{2(1-c)\sigma^2 T}   \right)   \;.
\end{equation}
We obtain the individual default probability by integrating this pdf from 0 up to the face value $F$,
\begin{eqnarray} \label{eq:PDXm}
\PD(X_{\rm m}) &=& \int\limits_0^{F} p_{V}(V(T))\, \rmd V(T) \nonumber \\
&=& \Phi\left( \frac{\ln\frac{F}{V_0}-\ln\left( X_{\rm m} + 1 \right) + \frac{1}{2}(1-c)\sigma^2 T}{\sqrt{(1-c)\sigma^2 T}} \right)  \;,
\end{eqnarray}
where $\Phi$ is the cumulative standard normal distrubution.
The expectation value for the individual loss given default $L^\gd=1-V(T)/F$ can be calculated as
\begin{eqnarray}
\langle L^\gd(X_{\rm m}) \rangle &=& \frac{1}{\PD(X_{\rm m})} \int\limits_0^{F} \left(1-\frac{V(T)}{F}\right) p_{V}(V(T))\, \rmd V(T)  \\
&=&  \frac{1}{\PD(X_{\rm m})} 
\left[  \Phi\left( \frac{\ln\frac{F}{V_0}-\ln\left( X_{\rm m} + 1 \right) + \frac{1}{2}(1-c)\sigma^2 T}{\sqrt{(1-c)\sigma^2 T}} \right) \right. \nonumber \\
&&
\left.  
-\exp\left( \ln\left( X_{\rm m} + 1 \right) - \ln\frac{F}{V_0}  \right)
\Phi\left( \frac{\ln\frac{F}{V_0}-\ln\left( X_{\rm m} + 1 \right) - \frac{1}{2}(1-c)\sigma^2 T}{\sqrt{(1-c)\sigma^2 T}} \right)
  \right] \;.   \nonumber
\end{eqnarray}
The expected recovery rate is simply
\begin{equation}
\langle R(X_{\rm m}) \rangle = 1-\langle L^\gd(X_{\rm m}) \rangle \;.
\end{equation}
And the portfolio loss for a homogeneous portfolio, or simply the average loss, is obtained as
\begin{equation} \label{eq:plossrec1}
\left< L(X_{\rm m}) \right> 
= \PD(X_{\rm m}) \langle L^\gd(X_{\rm m}) \rangle \;.
\end{equation}
For the sake of clarity, we introduce the function 
\begin{equation}
A(X_{\rm m}) = \ln\frac{F}{V_0}-\ln\left( X_{\rm m} + 1 \right)
\end{equation}
and the compound parameter 
\begin{equation}
B = \sqrt{(1-c)\sigma^2 T} \;.
\end{equation}
Now, the expressions for $\PD(X_{\rm m})$ and $\langle R(X_{\rm m}) \rangle$ simplify to
\begin{equation}\label{eq:PDAXm}
\PD(X_{\rm m}) = \Phi\left( \frac{ A(X_{\rm m}) + \frac{1}{2}B^2}{B} \right) 
\end{equation}
and 
\begin{equation}
\langle R(X_{\rm m}) \rangle = \rme^{-A(X_{\rm m})}\,
\Phi\left( \frac{A(X_{\rm m})-\frac{1}{2}B^2}{B} \right)
\left/ \Phi\left( \frac{A(X_{\rm m})+\frac{1}{2}B^2}{B} \right) \right.  \;.
\end{equation}
The relation between default probability and expected recovery rate depends only on $B$ and is parametrized by $A(X_{\rm m})$. Thus, the parameter $B$ can be calibrated to default and recovery rate data.
Furthermore, by inverting expression (\ref{eq:PDAXm}), we can express $A$ in terms of $\PD$,
\begin{equation}
A = B\, \Phi^{-1}(\PD)-\frac{1}{2}B^2 \;.
\end{equation}
This leads us to the functional dependence of recovery rate and default probability,
\begin{equation}  \label{eq:RPDrelation}
\langle R (\PD) \rangle = 
\frac{1}{\PD} \exp\left( -B\, \Phi^{-1}(\PD)+\frac{1}{2}B^2 \right)
\Phi\left( \Phi^{-1}(\PD) - B \right)  \;.
\end{equation}
It is worth repeating that this functional relation depends on a single parameter $B$ only.
In Figure~\ref{fig:b-dependence} we demonstrate this parameter dependence of Equation~(\ref{eq:RPDrelation}).
We plot $\langle R(\PD) \rangle$ for a set of parameter values. Larger values for $B$ lead to an overall decrease of recovery rates. In addition the dependence on $\PD$ becomes steeper.
From Equation~(\ref{eq:RPDrelation}) we obtain the functional relation of portfolio losses and default probabilities,
\begin{equation}  \label{eq:LPDrelation}
\langle L (\PD) \rangle = 
\PD-\exp\left( -B\, \Phi^{-1}(\PD)+\frac{1}{2}B^2 \right)
\Phi\left( \Phi^{-1}(\PD) - B \right)  \;.
\end{equation}
The parameter dependence of this relation is also shown in Figure~\ref{fig:b-dependence}.

In order to calculate Value at Risk and Expected Tail Loss we make use of the substitution
\begin{equation}
\left| p_L(L) \rmd L \right| = \left| p_{X_\m}(X_\m) \rmd X_\m \right| \;.
\end{equation}
Here we work with a simplified notation where $L$ refers to $\left< L(X_{\rm m}) \right>$.
With the cumulative distribution function
\begin{equation}
F_{X_{\m}}(x)=\int\limits_{-\infty}^{x} p_{X_{\m}}(X_{\m}) \rmd X_{\m}
\end{equation}
we can express the Value at Risk $\mathrm{VaR}_{\alpha}$ as
\begin{equation}
\mathrm{VaR}_{\alpha}=L(F_{X_{\m}}^{-1}(1-\alpha)) \;,
\end{equation}
where the function $L$ refers to the $X_\m$-dependence given in Equation~(\ref{eq:plossrec1}) and
the value $F_{X_{\m}}^{-1}(1-\alpha)$ is the $(1-\alpha)$-quantile of the market return. 
The Expected Tail Loss is calculated as
\begin{equation}
{\rm ETL}_\alpha = \frac{1}{\alpha} \int\limits_{{\rm VaR}_\alpha}^1 L\, p_L(L) \, \rmd L  
= \frac{1}{\alpha} \int\limits_{-\infty}^{F_{X_\m}^{-1}(1-\alpha)} L(X_\m)\, p_{X_\m}(X_\m) \, \rmd X_\m  \;.
\end{equation}

We will now compare these analytic results with Monte Carlo simulations for a finite portfolio size.

\subsection{Monte Carlo results}

In the Monte Carlo simulation we consider the stochastic process in Equation~(\ref{eq:VkTdiff})
for discrete time increments. 
The simulation is run with an inner loop and an outer loop.
In the inner loop we simulate $K=500$ different realizations of $\varepsilon_{k,t}$ for a single realization of the market fluctuations $\eta_{{\rm m},t}$  with $t=1,\ldots,N$. 
The inner loop can be interpreted as a homogeneous portfolio of size $K$, or simply as an average over the idiosyncratic part of the process. 
In each run of the inner loop, we calculate the market return $X_{\rm m}$, the number of defaults $N_\D(X_{\rm m})$ and the expected recovery rate $\left<R(X_{\rm m})\right>$.
The market return $X_{\rm m}$ is defined as the average return at maturity,
\begin{equation}
X_{\rm m} = \frac{1}{K}\sum\limits_{k=1}^K \left( \frac{V_k(T)}{V_0}-1 \right)  \;.
\end{equation}
For sufficiently large $K$ the idiosyncratic terms average out and the market return $X_{\rm m}$ is solely defined by the realization of $\eta_{{\rm m},t}$. This is the reason why we use the market return as a parameter for the other observables.
The number of defaults $N_\D(X_{\rm m})$ simply counts how many times the condition $V_k(T)<F$ is fulfilled. We can estimate the default probability as 
\begin{equation}\label{eq:pd}
\PD(X_{\rm m})\approx N_\D(X_{\rm m})/K \;.
\end{equation}
We obtain the portfolio loss as the average of individual losses in Equation~(\ref{eq:lossk}),   
\begin{equation}\label{eq:ploss}
\left< L(X_{\rm m}) \right> = \frac{1}{K} \sum_{k=1}^{K} L_k \;.
\end{equation}
Using the relation 
\begin{equation}\label{eq:plossrec}
\left< L(X_{\rm m}) \right> = \PD(X_{\rm m}) \left( 1-\left<R(X_{\rm m})\right> \right) 
\end{equation}
we can estimate the expected recovery rate as
\begin{equation}\label{eq:expectedR}
\left< R(X_{\rm m}) \right> 
= 1- \frac{\left<L(X_{\rm m})\right>}{\PD(X_{\rm m})} 
\approx 1- \frac{K \left<L(X_{\rm m})\right>}{N_\D(X_{\rm m})} \;.
\end{equation}
Here, we assume that the number of defaults is strictly non--zero, which is justified for large portfolio size $K$.
The outer loop runs over $10^6$ realizations of the market terms, where we obtain different values for the market return $X_{\rm m}$ and, consequently, the number of defaults $N_\D(X_{\rm m})$, the default probability $\PD(X_{\rm m})$ and the expected recovery rate $\left<R(X_{\rm m})\right>$.

In this paper we do not discuss the parameter dependence of the models or aim at calibrating them to a given portfolio. 
Instead we only present the results for a single set of parameters with economically sensible values.
As correlation coefficient we choose $c=0.5$. The parameters for the diffusion process are $\mu=0.05$ and $\sigma=0.15$. The initial market value is set to $V_0=100$, the face value of the zero--coupon bonds is $F=75$ with maturity time $T=1$.

\begin{figure*}
\begin{center}
 \incfig{0.49\textwidth}{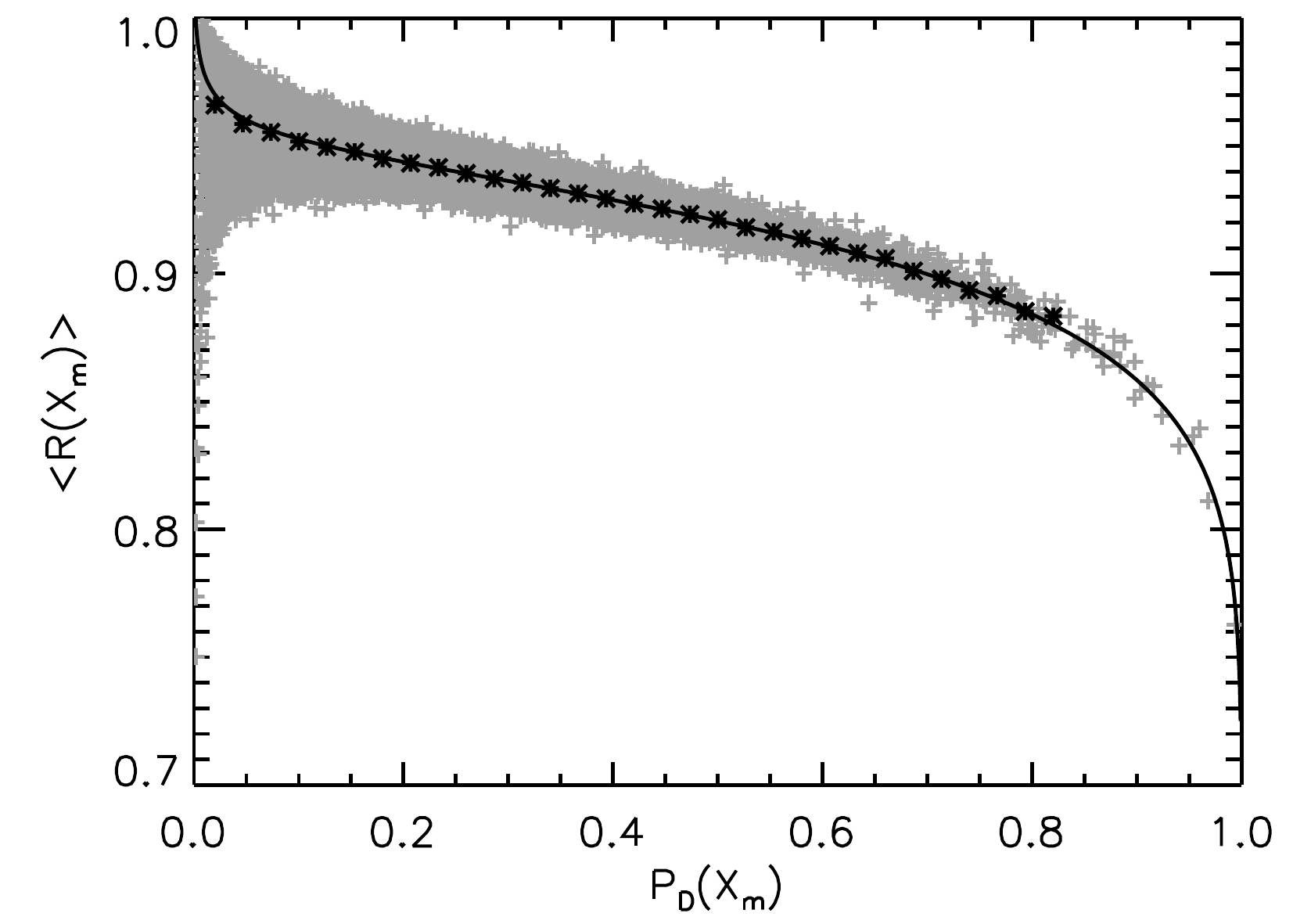}
 \incfig{0.49\textwidth}{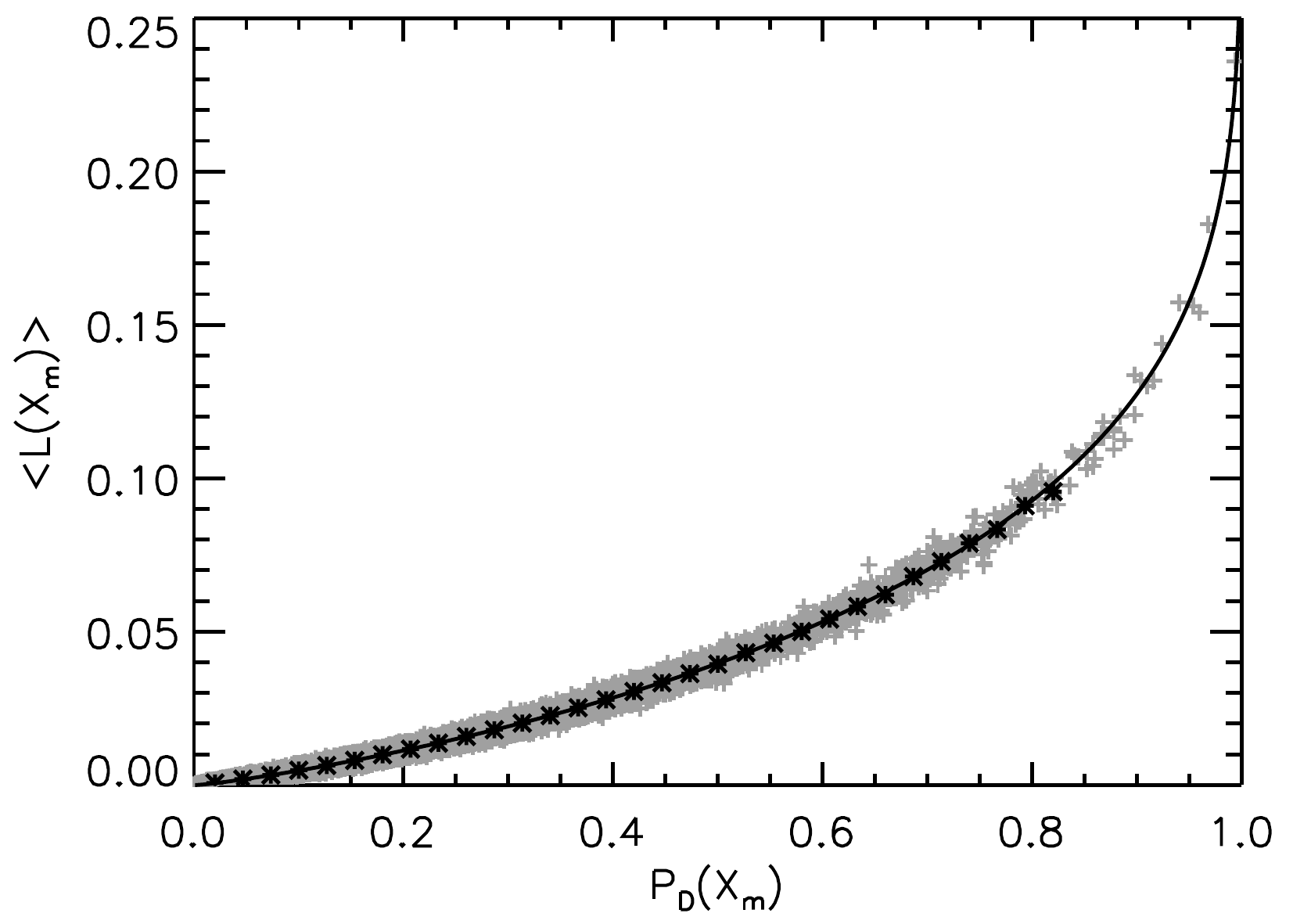}
\caption{
Dependence on default probabilities $\PD$ for the diffusion process.
The left plot shows the dependence of recovery rates $\langle R(X_\m) \rangle$ on default probabilities $\PD(X_\m)$. 
The right plot shows the dependence of portfolio losses $\langle L(X_\m) \rangle$ on default probabilities $\PD(X_\m)$.
The Monte Carlo results for the diffusion process (grey symbols) are compared to the analytical results (solid line). Black star symbols indicate local average values of the simulation results. The parameter $B=\sqrt{(1-c)\sigma^2 T}$ is determined from the simulation parameters.
}
\label{fig:pddep_diff}
\end{center}
\end{figure*}

\begin{figure*}
\begin{center}
 \incfig{0.49\textwidth}{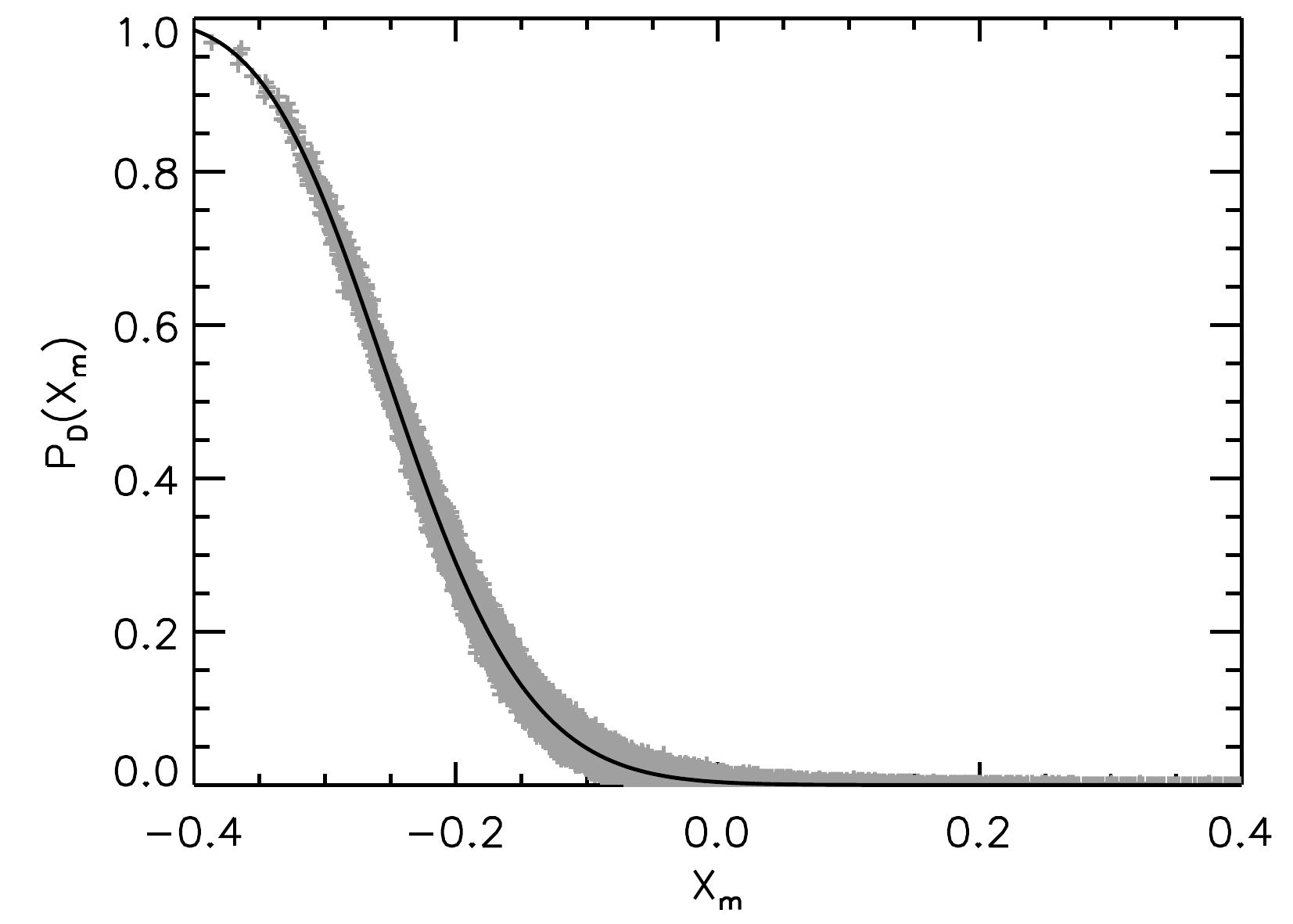}
 \incfig{0.49\textwidth}{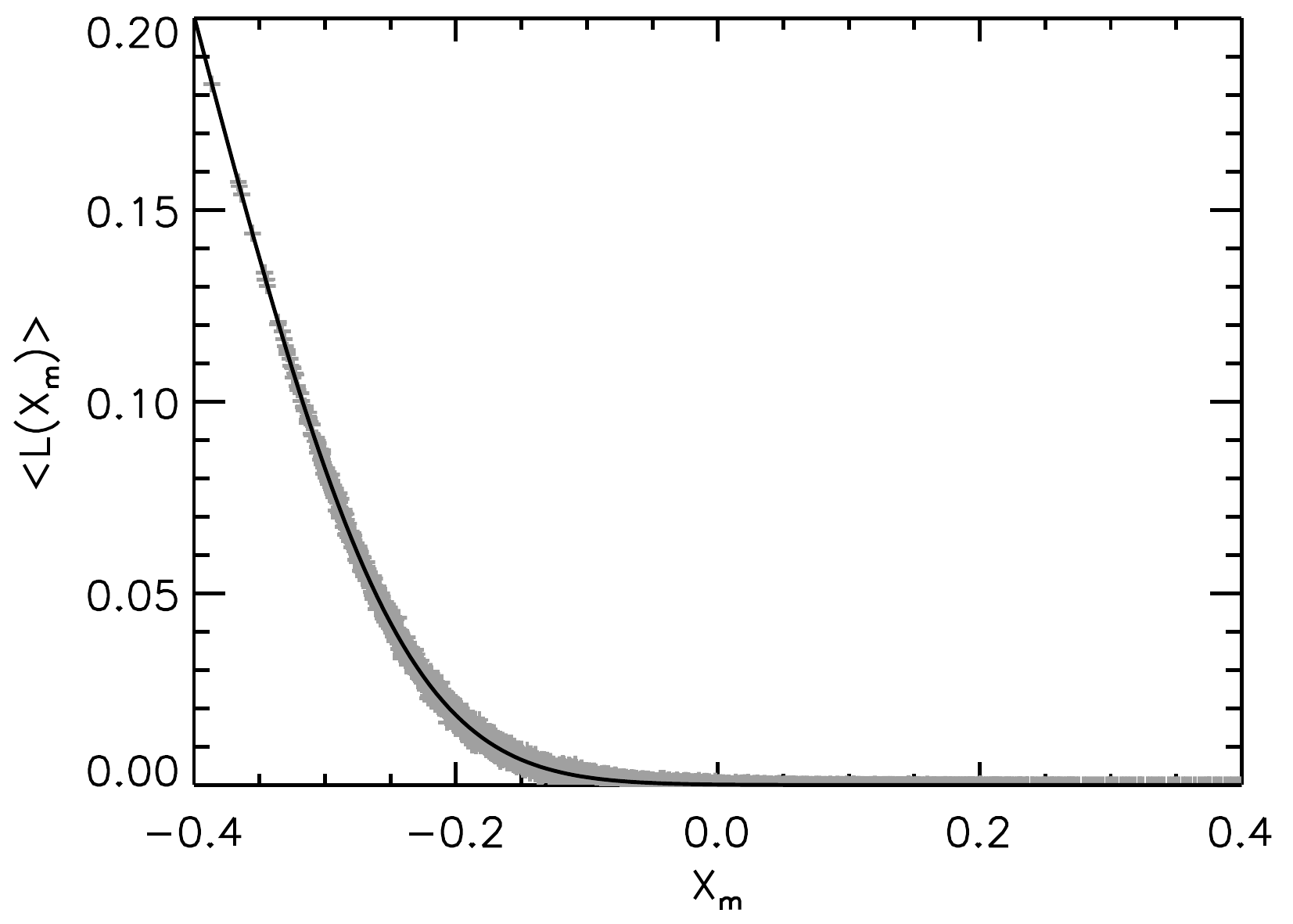}
\caption{
Dependence on market returns $X_\m$ for the diffusion process.
The left plot shows the dependence of default probabilities $\PD(X_\m)$ on market returns $X_\m$. 
The right plot shows the dependence of portfolio losses $\langle L(X_\m) \rangle$ on market returns $X_\m$.
The Monte Carlo results for the diffusion process (grey symbols) are compared to the analytical results (solid line). The parameter $B=\sqrt{(1-c)\sigma^2 T}$ is determined from the simulation parameters.
}
\label{fig:xmdep_diff}
\end{center}
\end{figure*}

Figure~\ref{fig:pddep_diff} shows the dependence of recovery rates $\langle R(X_\m) \rangle$ on default probabilities $\PD(X_\m)$ in one plot, and the dependence of portfolio losses $\langle L(X_\m) \rangle$ on default probabilities $\PD(X_\m)$ in a second plot.
In both cases we observe a very good agreement between the Monte Carlo simulations and the analytical results in Equations (\ref{eq:RPDrelation}) and (\ref{eq:LPDrelation}), respectively. Deviations from the average values are smaller for higher default probabilities.
Low default probabilities  correspond to positive or small negative market returns $X_\m$, see Figure~\ref{fig:xmdep_diff}. In this case defaults and recoveries are mostly influenced by the idiosyncratic part of the process and the correlation to the market plays only a minor role. This is why we observe a much broader range of recovery rates for low default probabilities.

When only limited historical data on defaults and recoveries is available, it might appear reasonable to assume a constant recovery rate.
This corresponds to a linear dependence of the portfolio loss on default probability. For small default probabilities this is a good approximation of the results shown in Figure~\ref{fig:pddep_diff}.
However, for larger default probabilities a constant recovery rate model severly underestimates portfolio losses.
The combination of high default probabilities and low recovery rates strongy influences the tail of the loss distribution. 

\begin{figure*}
\begin{center}
 \incfig{0.49\textwidth}{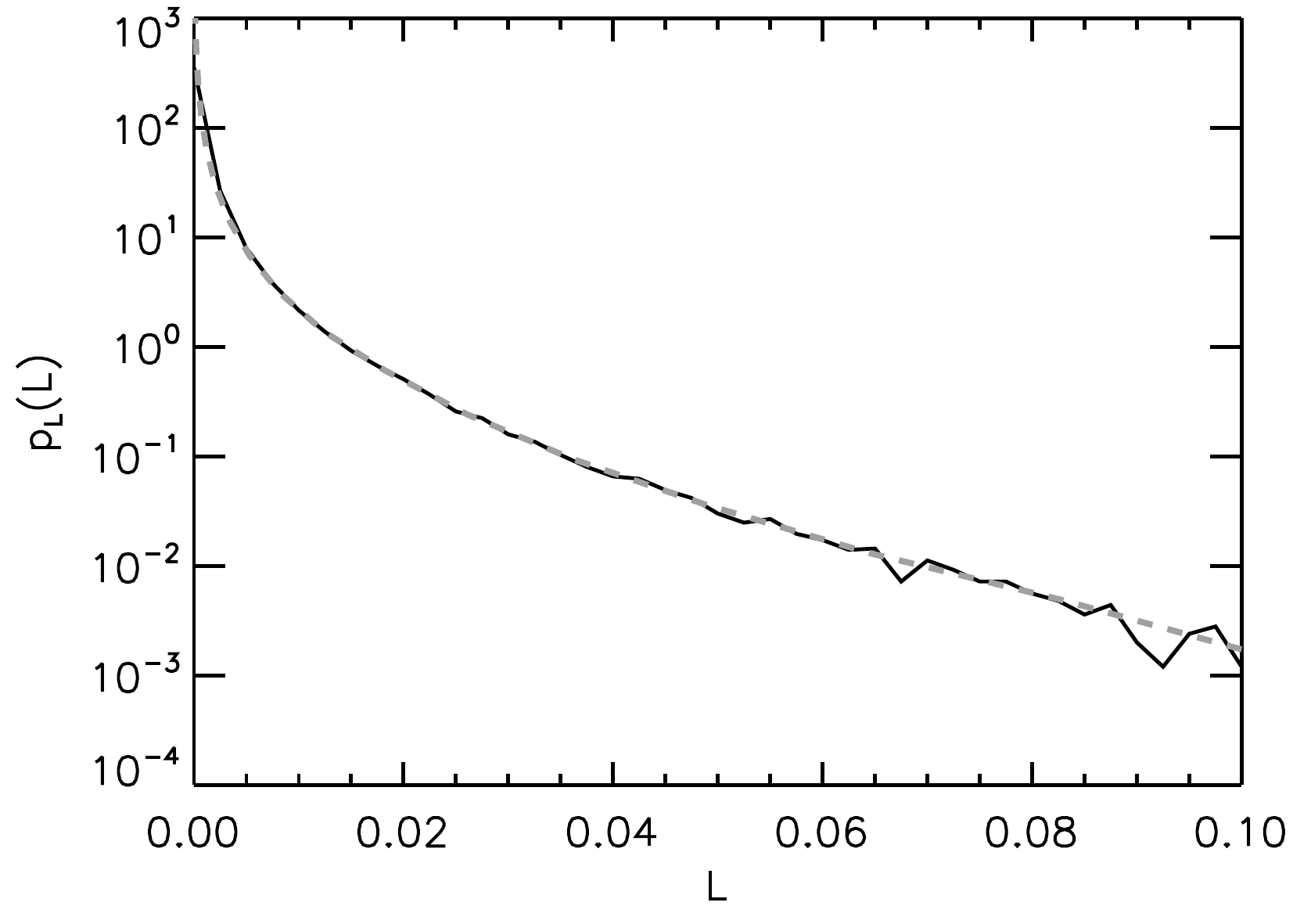}
\caption{
Probability density function $p_L(L)$ for the portfolio losses for the diffusion process.
The black solid line corresponds to the simulation results.
The grey dashed line shows the analytical results, where the pdf of market returns is transformed.
}
\label{fig:ploss_diff}
\end{center}
\end{figure*}

In Figure~\ref{fig:xmdep_diff} we demonstrate that the dependence of default probabilities and portfolio losses on the market return is well described by Equations (\ref{eq:PDXm}) and (\ref{eq:plossrec1}), respectively. Given the functional relation between portfolio loss $L$ and market return $X_\m$ we can transform the pdf of market returns into the pdf of portfolio losses,
\begin{equation}  \label{eq:pLpXm}
p_L(L)=\frac{1}{\left| L^\prime(X_\m)  \right|} \, p_{X_\m}(X_\m)   \;.
\end{equation}
A comparison to the Monte Carlo result for the loss distribution is given in Figure~\ref{fig:ploss_diff}. We observe an excellent agreement, even for extremely large portfolio losses.
Finally, we present the results for Expected Loss, Value at Risk and Expected Tail Loss in Table~\ref{tab:diff}. The latter two are calculated for the 0.99--quantile. The analytical values describe the simulation results very well. 

Since the diffusion process was used both in the Monte Carlo simulations and in the analytical derivation, the good agreement of the respective results comes as no surprise. In the following sections we will explore two different processes in the Monte Carlo simulations and compare these with the analytical results for the diffusion.

\begin{table}[htdp]
\begin{center}
\begin{tabular}{l c c} \hline
& \ simulation \ & \ analytical result   \\ \hline
EL  \ &  $7.38 \cdot 10^{-4}$  &  $7.35 \cdot 10^{-4}$     \\
VaR$_{0.99}$ \ &  $1.30 \cdot 10^{-2}$  &     $1.29 \cdot 10^{-2}$    \\
ETL$_{0.99}$ \ &    $2.38 \cdot 10^{-2}$   &   $2.37 \cdot 10^{-2}$ \\ \hline 
\end{tabular} \caption{Expected Loss, Value at Risk and Expected Tail Loss for the diffusion process.}
\end{center}
\label{tab:diff}
\end{table}

\section{Correlated jump--diffusion} \label{sec4}

\subsection{Analytical discussion}

We extend the diffusion model of the previous section by adding two jump terms to the stochastic process, $\rmd J_k$ for idiosyncratic jumps and $\rmd J_ {\rm m}$ for jumps which affect the entire market. The jump terms are not contained in Merton's original model; they ensure that the default probability does not vanish as the time to maturity becomes very short.
For applications of jump--diffusion processes in credit risk modelling see, e.g., \cite{zhou01},
\cite{schaefer07a} and \cite{KieselScherer2007}.
The stochastic differential equation for the market value reads
\begin{equation} \label{eq:jumpdiff}
\frac{\rmd V_k}{V_k} = \mu \rmd t +  \sqrt{c}\, \sigma \, \rmd W_{\rm m}  + \rmd J_ {\rm m}
 + \sqrt{1-c}\, \sigma \, \rmd W_k  + \rmd J_k \ .
\end{equation}
This is a correlated jump--diffusion process.
We model the jumps by a {\it Poisson process} with intensity $\lambda$. We
recall that in such a process the probability function for the event
to occur $n$ times between zero and time $t$ is given by
\begin{equation}\label{pp}
p_n^{{\rm Poisson}}(t) = 
  \frac{(\lambda t)^n}{n!}\exp\left(-\lambda t\right) \ .
\end{equation}

The size $\Lambda$ of the jump, measured in units of the current market
value $V_k(t)$, is a random variable with a distribution which we have
to specify.  Jumps can be positive or negative. The largest possible
negative jump is 100\% of the current market value. Based on this
information, a possible distribution of the jump size $\Lambda$ is a
shifted lognormal distribution, 
$\Lambda + 1 \sim {\rm LN}(\mu_J,\sigma_J)$, with mean $\mu_J$ and standard deviation
$\sigma_J$.

Without the jump term, the distribution of the market value $V_k(t)$ is
log--normal. The jumps render the tails of the distribution fatter. 
The parameters of the jump process can be adjusted in order to match the tail behavior of a given empirical time series of the market value. Here, we use the same parameters for idiosyncratic and market wide jumps.   

Let us now consider the stochastic process in Eq.~(\ref{eq:jumpdiff}) for discrete time increments $\Delta t=T/N$, where the time to maturity $T$ is divided into $N$ steps. Then the market value of company $k$ at maturity is given by
\begin{equation}\label{eq:VkT}
V_k(T)=V_0 \prod\limits_{t=1}^{N} 
\left(1+ \mu \Delta t +  \sqrt{c}\, \sigma \eta_{{\rm m},t} \sqrt{\Delta t} + \rmd J_{{\rm m},t}
 + \sqrt{1-c}\, \sigma \varepsilon_{k,t} \sqrt{\Delta t} + \rmd J_{k,t}\right) \;.
\end{equation}
Akin to the diffusion case, we find for the market return of the jump--diffusion
\begin{equation}
X_{\rm m} = \frac{1}{K}\sum\limits_{k=1}^K \prod\limits_{t=1}^{N} 
\left(1+ \mu \Delta t +  \sqrt{c}\, \sigma \eta_{{\rm m},t} \sqrt{\Delta t} + \rmd J_{{\rm m},t}
 + \sqrt{1-c}\, \sigma \varepsilon_{k,t} \sqrt{\Delta t} + \rmd J_{k,t}\right) -1 \;.
\end{equation}
For $K\to\infty$ we can express the average over $k$ as expectation values for $\varepsilon_{k,t}$ and $\rmd J_{k,t}$. Since the averages are independent for different $k$ and $t$, we can write
\begin{equation}\label{eq:Xmjd}
X_{\rm m} + 1 =  \prod\limits_{t=1}^{N} 
\left(1+ \mu \Delta t +  \sqrt{c}\, \sigma \eta_{{\rm m},t} \sqrt{\Delta t} + \rmd J_{{\rm m},t}
 + \sqrt{1-c}\, \sigma \langle \varepsilon_{k,t} \rangle \sqrt{\Delta t} + \langle \rmd J_{k,t} \rangle \right) \;,
\end{equation}
where the averages are given as
\begin{eqnarray}
\langle \varepsilon_{k,t} \rangle &=& 0  \\
\langle \rmd J_{k,t} \rangle &=& \left(\exp\left( \mu_J+ \frac{\sigma_J^2}{2} \right) -1 \right)\lambda_{\rm J} \;.
\end{eqnarray}
Unless we choose $\mu_J=- \sigma_J^2/2$, the expectation value for the jump term does not vanish.
In the diffusion case we expanded the logarithm in Equation~(\ref{eq:Xmapprox}). An analogous step is not possible for the jump--diffusion, because the jump term $\rmd J_{{\rm m},t}$ does not scale with $\Delta t$.
This complicates a further analytical discussion of the jump--diffusion case.

\subsection{Monte Carlo results}

In the Monte Carlo simulation we consider the stochastic process in Equation~(\ref{eq:VkT})
for discrete time increments.
As in the diffusion case we simulate $K=500$ different realizations of $\varepsilon_{k,t}$ and $\rmd J_{k,t}$ for a single realization of the market terms $\eta_{{\rm m},t}$ and $\rmd J_{{\rm m},t}$ with $t=1,\ldots,N$. 
For each realization of the idiosynchratic terms we calculate the market return $X_{\rm m}$, the number of defaults $N_\D(X_{\rm m})$, the default probability $\PD(X_{\rm m})$ and the expected recovery rate $\left<R(X_{\rm m})\right>$.
The full Monte Carlo simulation consists of $10^6$ realizations of the market terms.

For the diffusion terms and the contract details we use the same parameters as in Section~\ref{sec3}. The additional parameters for the jump terms are $\lambda=0.005$, $\mu_{\mathrm{j}}=0.4$ and $\sigma_{\mathrm{j}}=0.3$.

\begin{figure*}
\begin{center}
 \incfig{0.49\textwidth}{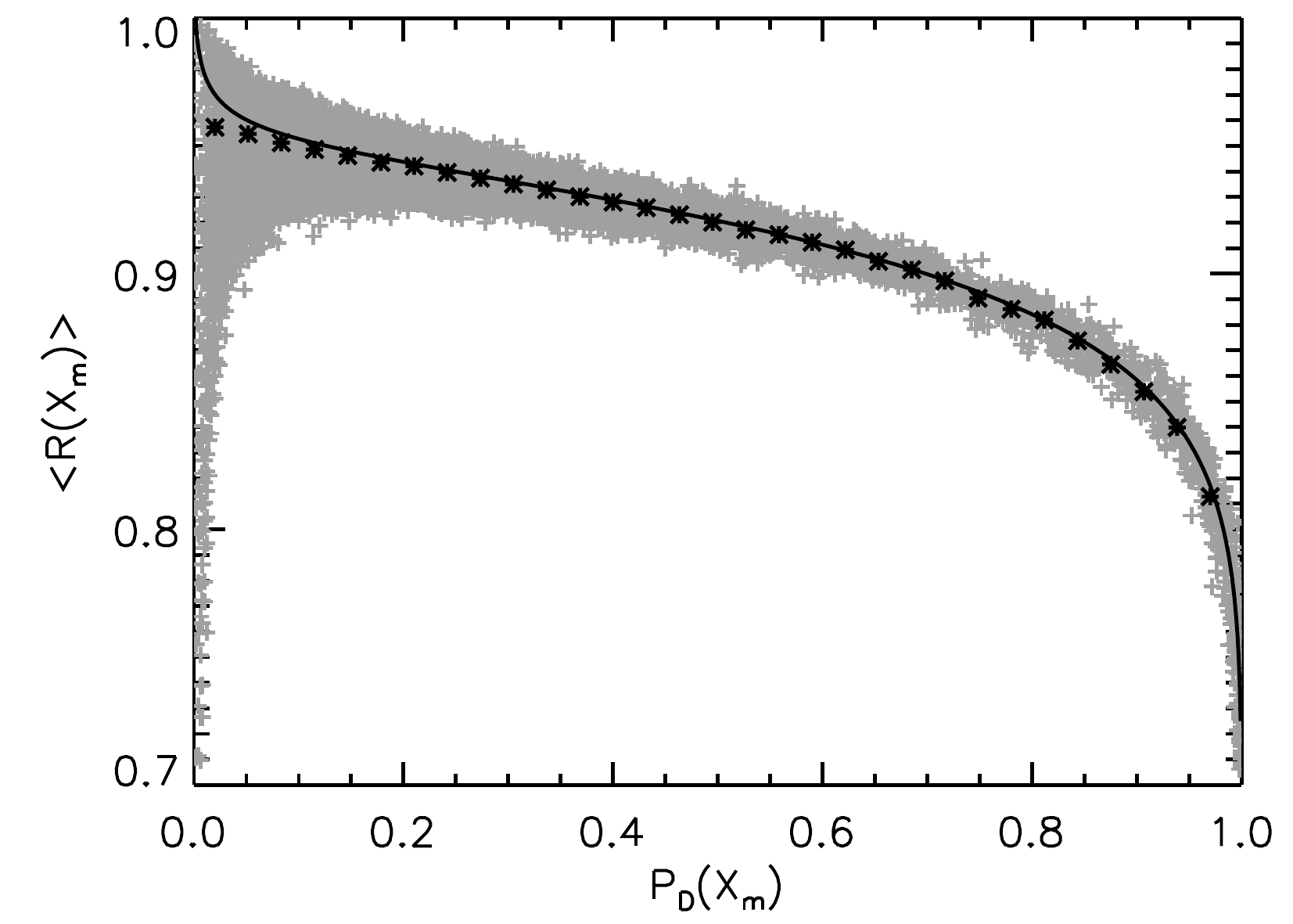}
 \incfig{0.49\textwidth}{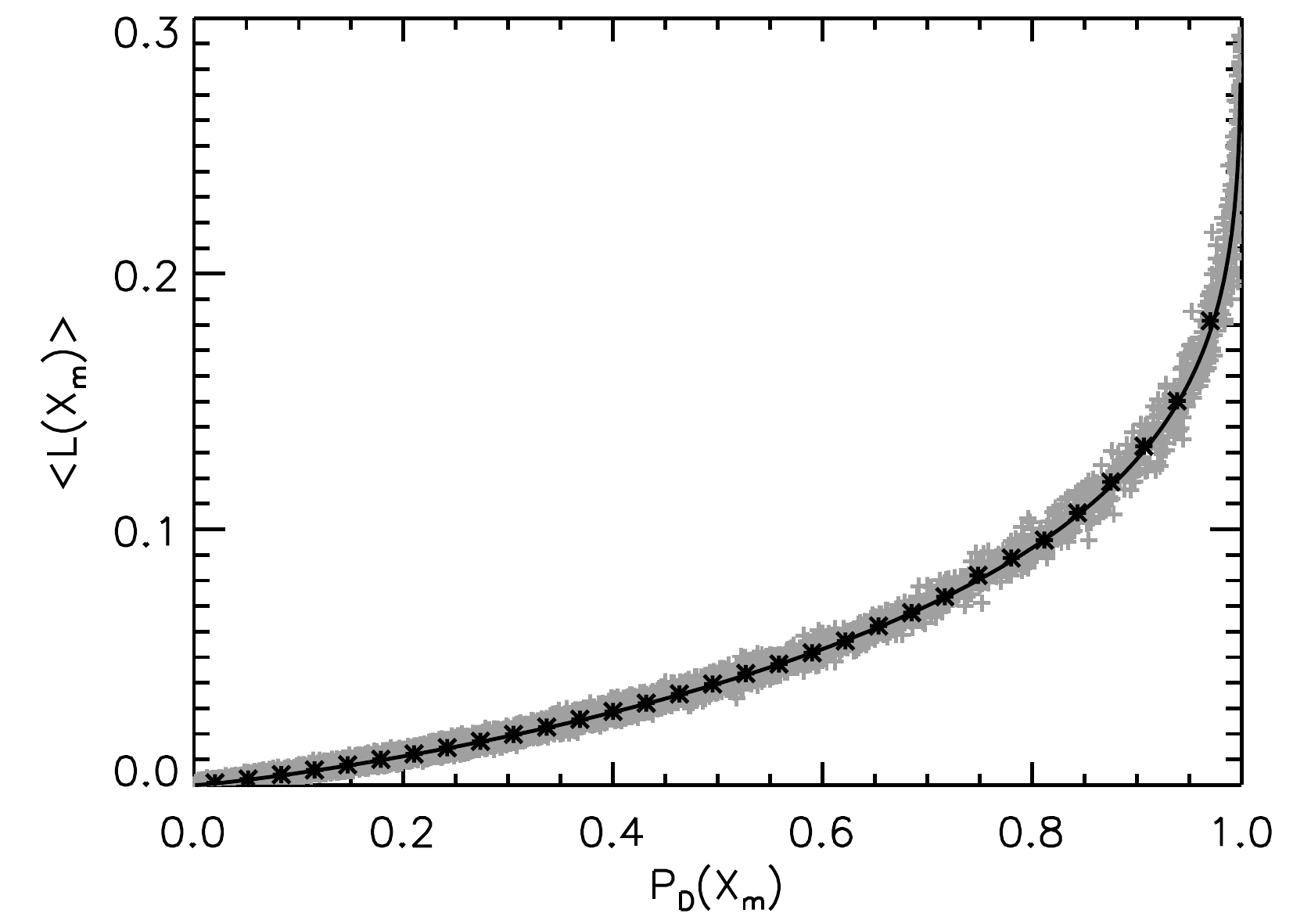}
\caption{
Dependence on default probabilities $\PD$ for the jump--diffusion process.
The left plot shows the dependence of recovery rates $\langle R(X_\m) \rangle$ on default probabilities $\PD(X_\m)$. 
The right plot shows the dependence of portfolio losses $\langle L(X_\m) \rangle$ on default probabilities $\PD(X_\m)$.
The Monte Carlo results for the jump--diffusion process (grey symbols) are compared to the analytical results for the diffusion (solid line). The parameter $B=\sqrt{(1-c)\sigma^2 T}$ is determined from the simulation parameters.
}
\label{fig:pddep_jdiff}
\end{center}
\end{figure*}

Figure~\ref{fig:pddep_jdiff} shows the dependence of recovery rates $\langle R(X_\m) \rangle$ on default probabilities $\PD(X_\m)$ in one plot, and the dependence of portfolio losses $\langle L(X_\m) \rangle$ on default probabilities $\PD(X_\m)$ in a second plot.
In both cases we observe a very good agreement between the Monte Carlo simulations of the jump--diffusion and the analytical results for the diffusion, see Equations (\ref{eq:RPDrelation}) and (\ref{eq:LPDrelation}), respectively.
The average recovery rates deviate only for very low default probabilities from the diffusion result. For the portfolio loss this deviation is suppressed and no longer visible.
The parameter $B$ is here the same as in the diffusion case, this means it is not influenced by the additional jump terms of the process.
The most striking difference to the plots in Figure~\ref{fig:pddep_diff} is the abundance of simulation results with very high default probabilities.
This is due to the market--wide jumps which render the tails of the market return distribution fatter.
The idiosyncratic jumps lead to slightly larger deviations of individual results from the mean values.

\begin{figure*}
\begin{center}
 \incfig{0.49\textwidth}{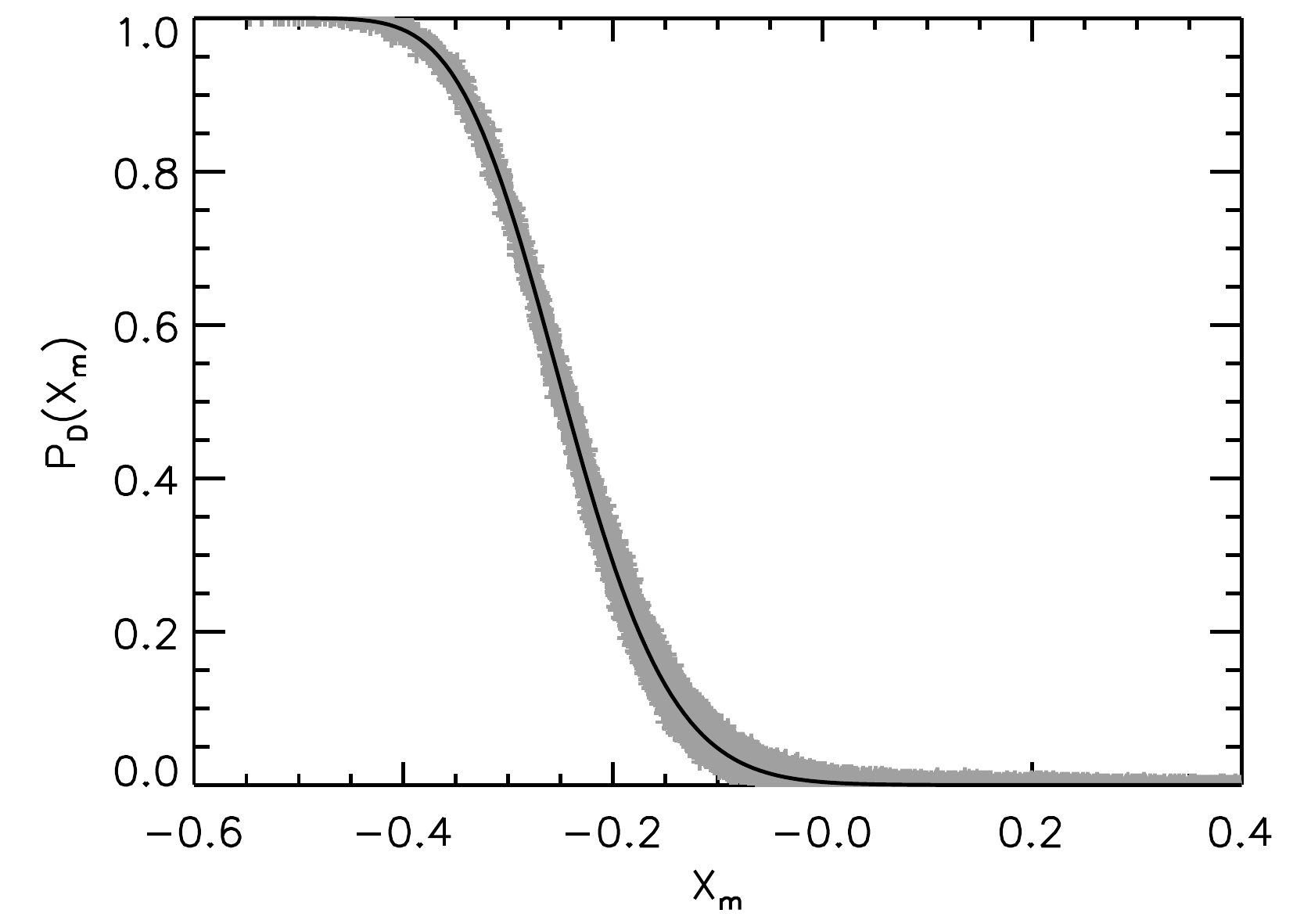}
 \incfig{0.49\textwidth}{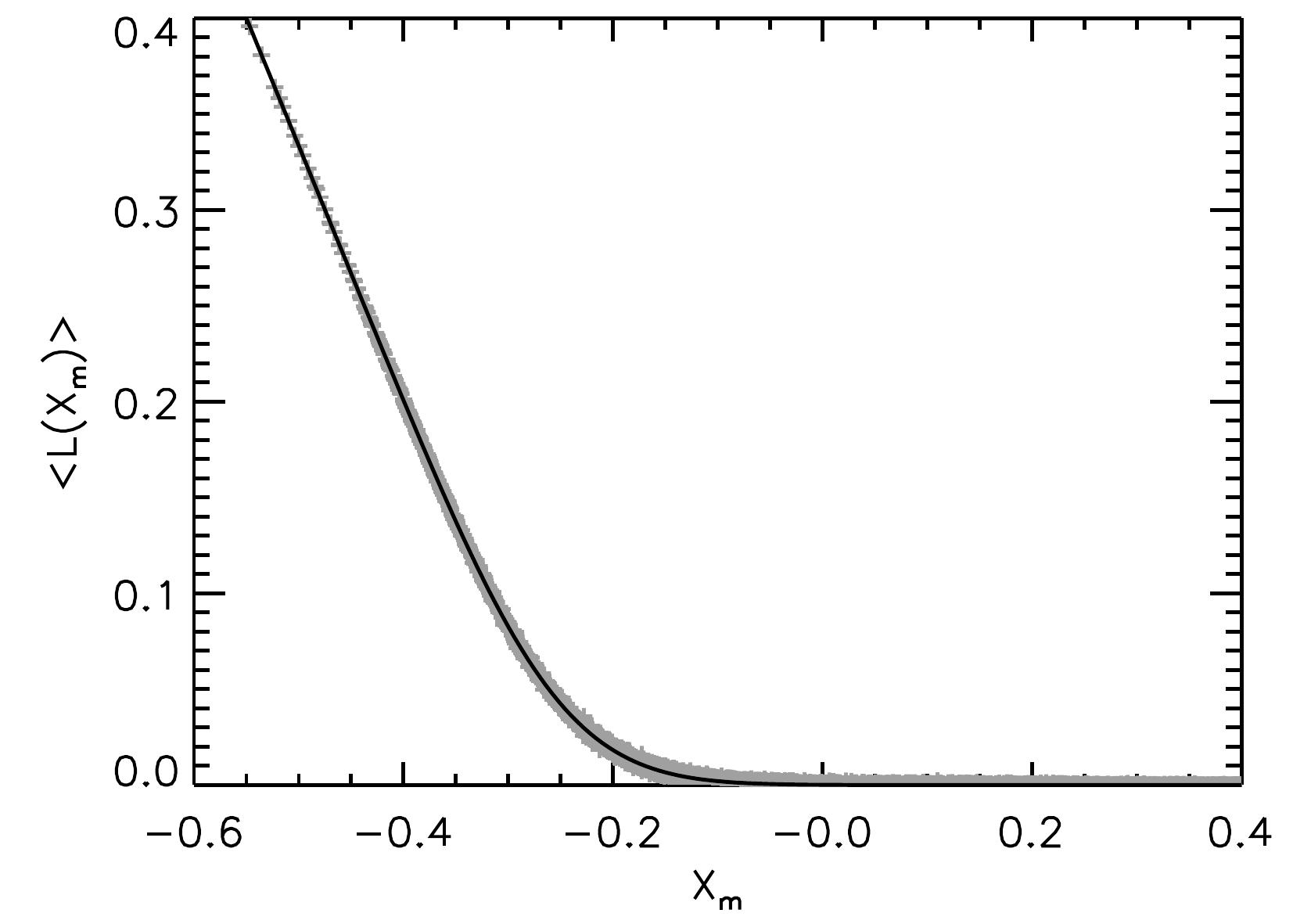}
\caption{
Dependence on market returns $X_\m$ for the jump--diffusion process.
The left plot shows the dependence of default probabilities $\PD(X_\m)$ on market returns $X_\m$. 
The right plot shows the dependence of portfolio losses $\langle L(X_\m) \rangle$ on market returns $X_\m$.
The Monte Carlo results for the jump--diffusion process (grey symbols) are compared to the analytical results for the diffusion (solid line). The parameter $B=\sqrt{(1-c)\sigma^2 T}$ is determined from the simulation parameters.
}
\label{fig:xmdep_jdiff}
\end{center}
\end{figure*}

\begin{figure*}
\begin{center}
 \incfig{0.49\textwidth}{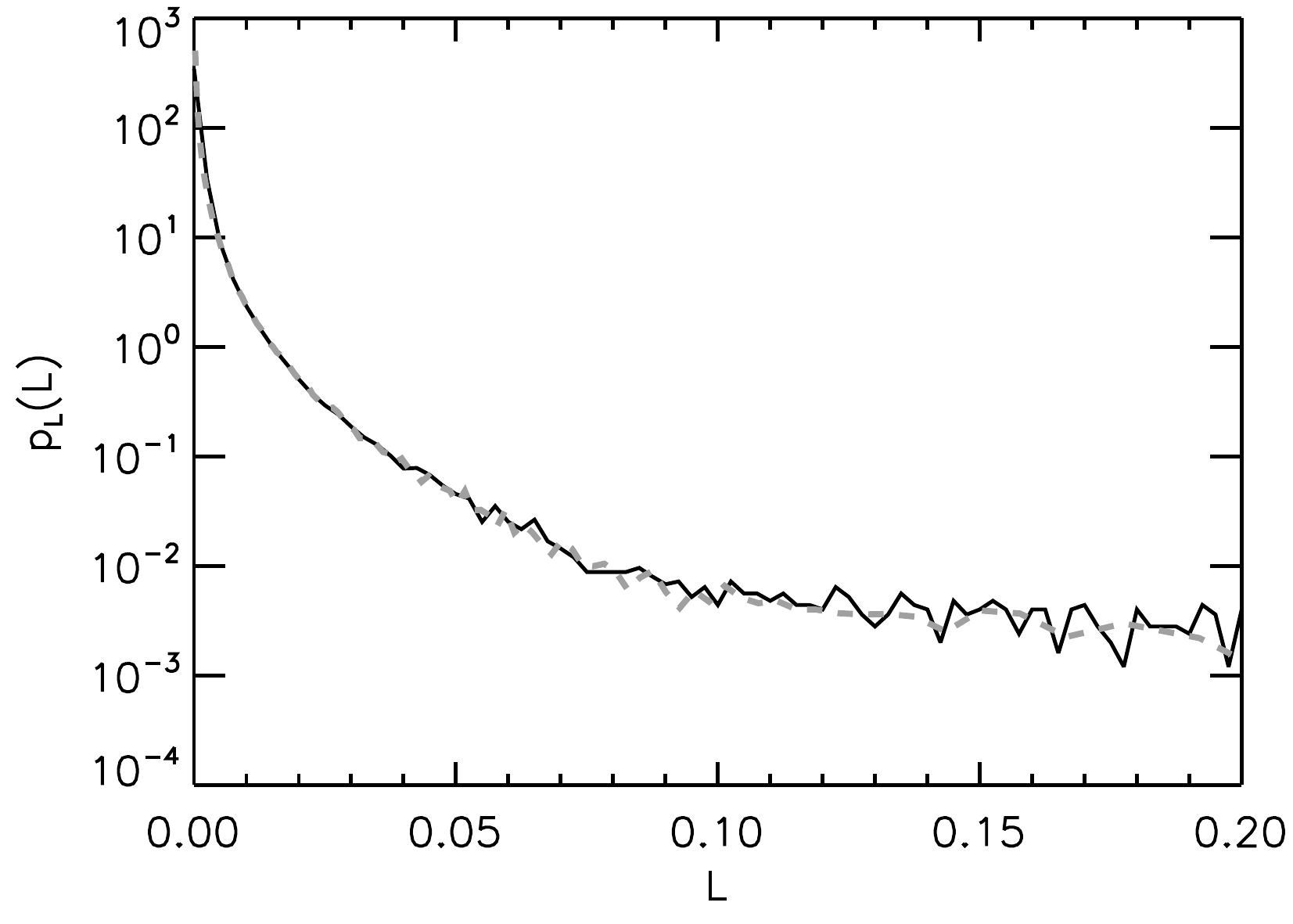}
\caption{
Probability density function $p_L(L)$ for the portfolio losses for the jump--diffusion process.
The black solid line corresponds to the simulation results.
The grey dashed line shows the analytical results, where the pdf of market returns is transformed.
}
\label{fig:ploss_jdiff}
\end{center}
\end{figure*}

The analytical results for the diffusion case do not only describe the $\PD$ dependence of the jump--diffusion results, but also their  dependence on the market return $X_\m$. This is shown in Figure~\ref{fig:xmdep_jdiff}.
Both default probabilities and portfolio losses are well described by Equations (\ref{eq:PDXm}) and (\ref{eq:plossrec1}), respectively. Given the functional relation between portfolio loss $L$ and market return $X_\m$ we can again use Equation~\ref{eq:pLpXm} to transform the pdf of market returns into the pdf of portfolio losses. Here we use the simulation data on $X_\m$ to determine their distribution numerically.
In Figure~\ref{fig:ploss_jdiff} we compare this to the Monte Carlo result for the loss distribution. As in the diffusion case, we observe an excellent agreement, even for extremely large losses. The market--wide jumps lead to a very slow decay in the tail of the loss distribution.
The results for Expected Loss, Value at Risk and Expected Tail Loss are listed in Table~\ref{tab:jdiff}. 
The analytical values describe the simulation results very well. 

Our results show that the Merton model with jump--diffusion can be described by the diffusion results if we exchange the distribution of market returns. The dependencies on default probabilities and market returns are nearly the same as in the diffusion case.
This is somewhat reminiscent of a copula approach to credit risk.  

\begin{table}[htdp]
\begin{center}\begin{tabular}{l c c} \hline
& \ simulation \ & \ analytical result  \\ \hline
EL \  &  $1.03 \cdot 10^{-3}$  &  $9.07 \cdot 10^{-4}$     \\
VaR$_{0.99}$ \ &  $1.50 \cdot 10^{-2}$  &     $1.47 \cdot 10^{-2}$   \\
ETL$_{0.99}$ \ &    $3.70 \cdot 10^{-2}$   &   $3.75 \cdot 10^{-2}$  \\ \hline
\end{tabular} \caption{Expected Loss, Value at Risk and Expected Tail Loss for the jump--diffusion process.}
\end{center}
\label{tab:jdiff}
\end{table}

\section{Correlated GARCH process} \label{sec5}

Finally, we examine the Merton model with a GARCH(1,1) model for the underlying asset value process. The GARCH model was first introduced by \cite{bollerslev86}. It is able to reproduce many of the stylized facts found in empirical financial time series. In particular, it exhibits volatility clustering and fat--tailed return distributions.
The GARCH(1,1) model is a discrete time process with an autoregressive volatility. The return at time $t$ reads
\begin{eqnarray}\label{eq:garch}
r_{k,t} & = & \sigma_{k,t}\, \left( \sqrt{c}\, \eta_t + \sqrt{1-c}\, \varepsilon_{k,t} \right)    \nonumber \\ 
\sigma_{k,t}^2 & = & \alpha_0 + \alpha_1\, r_{k,t-1}^2 + \beta_1\, \sigma_{k,t-1}^2
\;.
\end{eqnarray}
where $\eta_t$ and $\varepsilon_{k,t}$ are independent normal distributed random variables.
The parameters $\alpha_0$, $\alpha_1$ and $\beta_1$ are chosen in order to mimic the behavior of a typical empirical time series of daily returns. The initial values for the volatilities have been set homogeneously as $\sigma_{k,0}=\sigma\sqrt{\Delta t}$, where $\sigma=0.15$ is the same value used in the diffusion and the jump--diffusion case.
However, the volatility does not remain the same for all $k$ as $t$ evolves, since $\sigma_{k,t}$ also depends on the idiosyncratic random part in $r_{k,t-1}$. Thus, the homogeneity of the portfolio is lost to some degree. Instead the process covers a wide range of volatilities both within one realization of the market terms, i.e., within one portfolio, and also over different market realizations.
The GARCH process is therefore the most general case we examine here. It can provide a good indication of how broadly our analytical results for the diffusion case can be applied.

For the GARCH process the market value of company $k$ at maturity reads
\begin{equation}\label{eq:VkTgarch}
V_k(T)=V_0 \prod\limits_{t=1}^{N} 
\left(1+ \mu \Delta t + r_{k,t}  \right) \;.
\end{equation}
The deterministic drift term can also be included in Equation~(\ref{eq:garch}) instead.
Then the parameters  $\alpha_0$, $\alpha_1$ and $\beta_1$ have to be adjusted accordingly.
The drift constant is again set to $\mu=0.05$.

As in the diffusion and jump--diffusion case we simulate $10^6$ portfolios of size $K=500$. 
For each portfolio we calculate the market return $X_{\rm m}$, the number of defaults $N_\D(X_{\rm m})$, the default probabilities $\PD(X_\m)$ and the expected recovery rate $\left<R(X_{\rm m})\right>$.

\begin{figure*}
\begin{center}
 \incfig{0.49\textwidth}{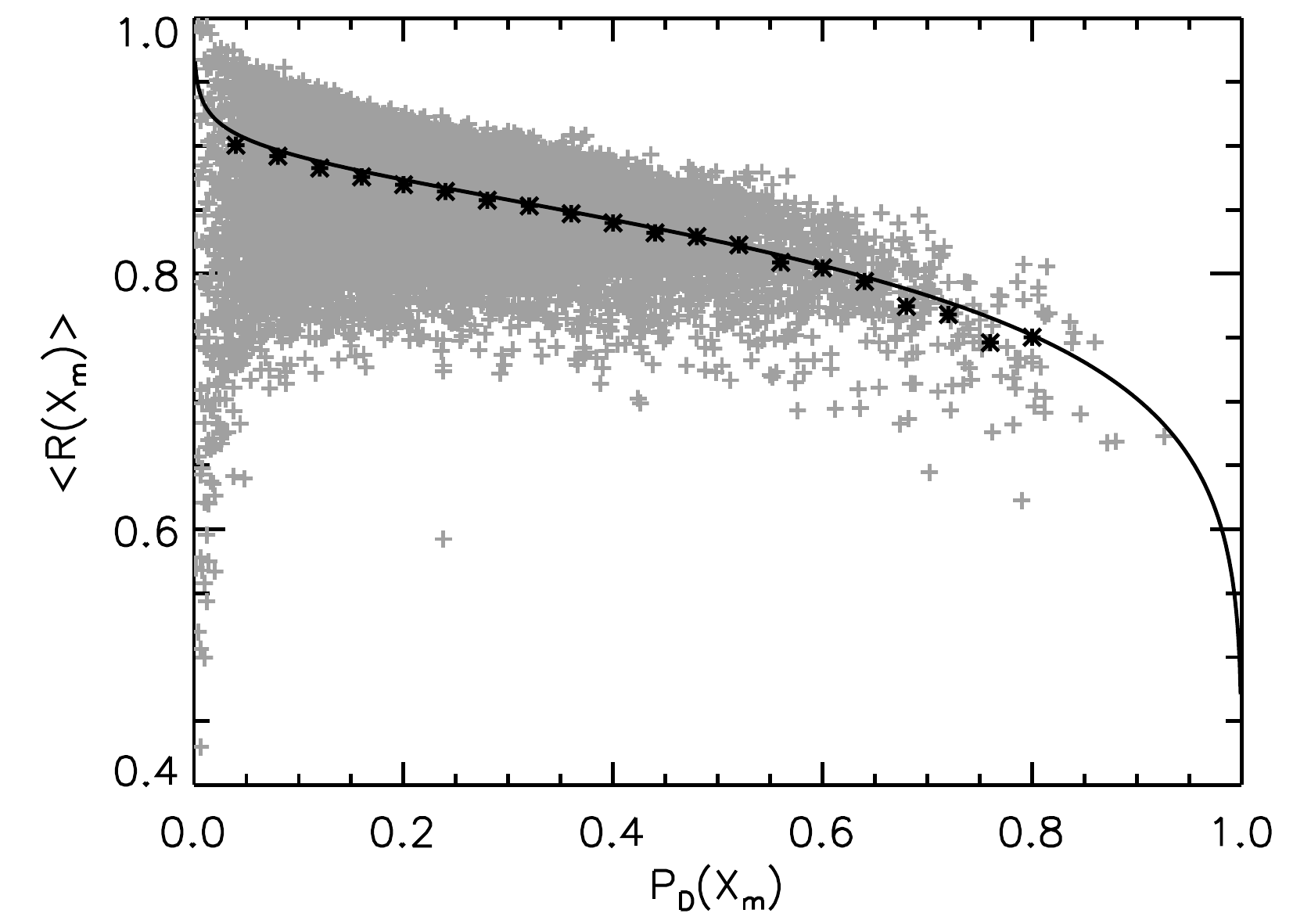}
 \incfig{0.49\textwidth}{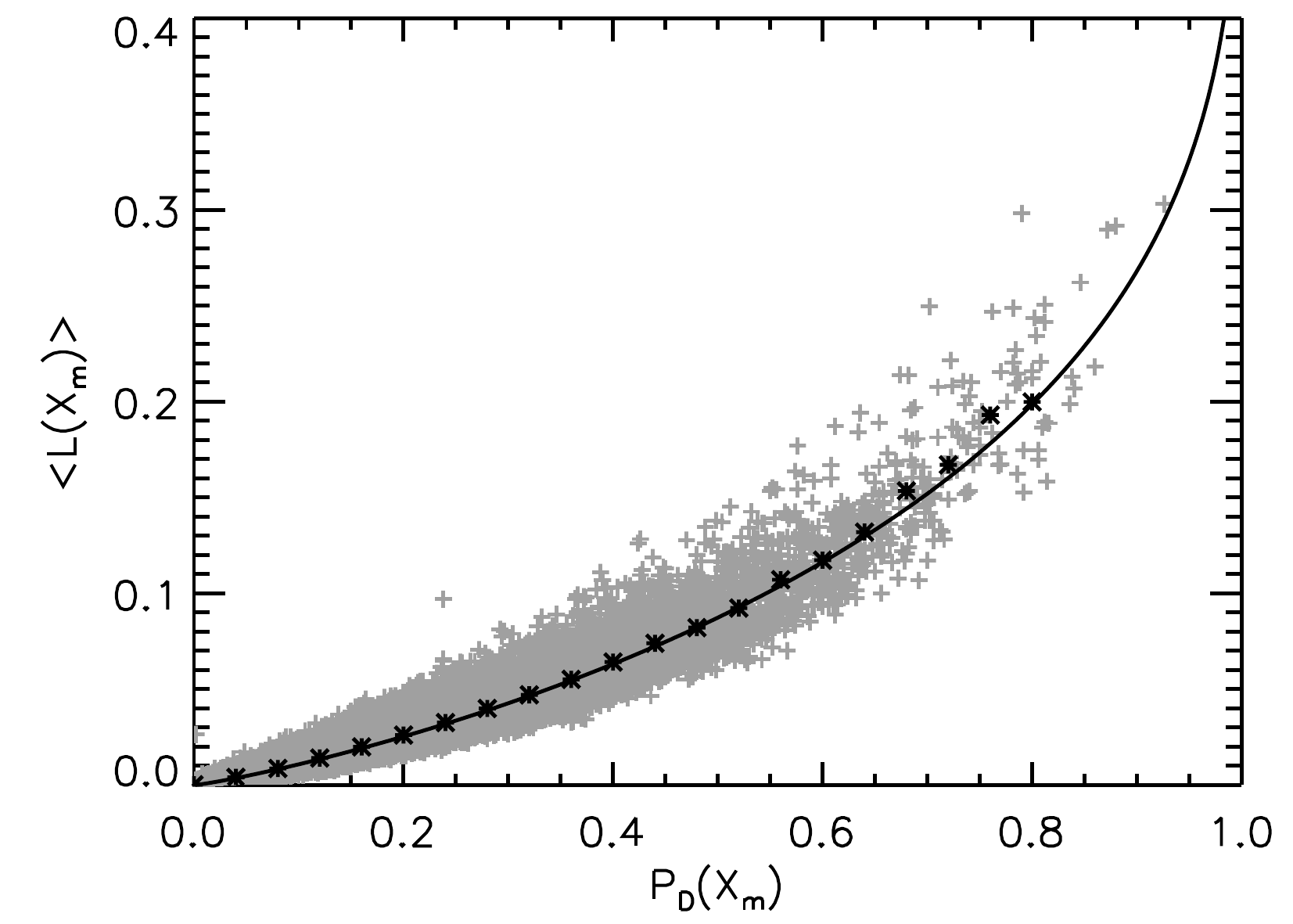}
\caption{
Dependence on default probabilities $\PD$ for the GARCH process.
The left plot shows the dependence of recovery rates $\langle R(X_\m) \rangle$ on default probabilities $\PD(X_\m)$. 
The right plot shows the dependence of portfolio losses $\langle L(X_\m) \rangle$ on default probabilities $\PD(X_\m)$.
The Monte Carlo results for the GARCH process (grey symbols) are compared to the analytical results for the diffusion (solid line). The parameter $B$ is fitted to the data.
}
\label{fig:pddep_garch}
\end{center}
\end{figure*}

In Figure~\ref{fig:pddep_garch} we present the Monte Carlo results for the $\PD$ dependence of the GARCH process. One plot shows the dependence of recovery rates $\langle R(X_\m) \rangle$ on default probabilities $\PD(X_\m)$, a second plot shows the dependence of portfolio losses $\langle L(X_\m) \rangle$ on default probabilities $\PD(X_\m)$.
First, we observe a much broader range of the individual results than in the diffusion and jump--diffusion case. This is to be expected because the fluctuating volatilities lead to pronounced fat tails in the return distribution of individual companies.
While the returns are correlated in the GARCH(1,1) model, volatilities are rather uncorrelated, see  \cite{SchaeferGuhr2010}.
Thus, for the market return the fluctuations of individual volatilities average out to some degree.
This is why we scarcely observe very high default probabilities in the simulations.
The situation is different for empirical stock return time series, where volatilities are also strongly correlated.

It is a rather remarkable result that the average behavior of the GARCH simulation is so well described by the analytical results for the diffusion, see Equations (\ref{eq:RPDrelation}) and (\ref{eq:LPDrelation}), respectively.
The parameter $B$ has been fitted to the simulation data.

\begin{figure*}
\begin{center}
 \incfig{0.49\textwidth}{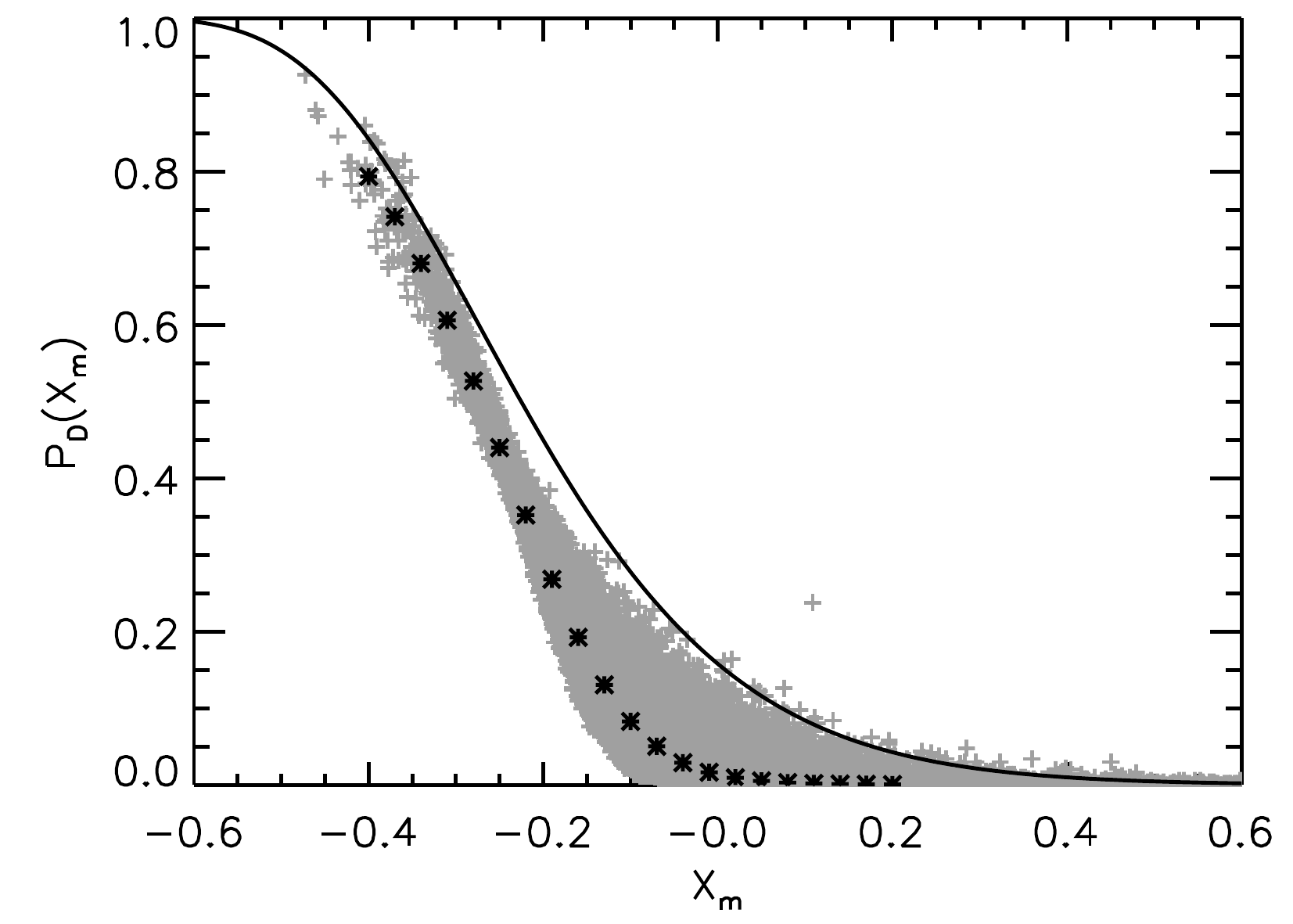}
 \incfig{0.49\textwidth}{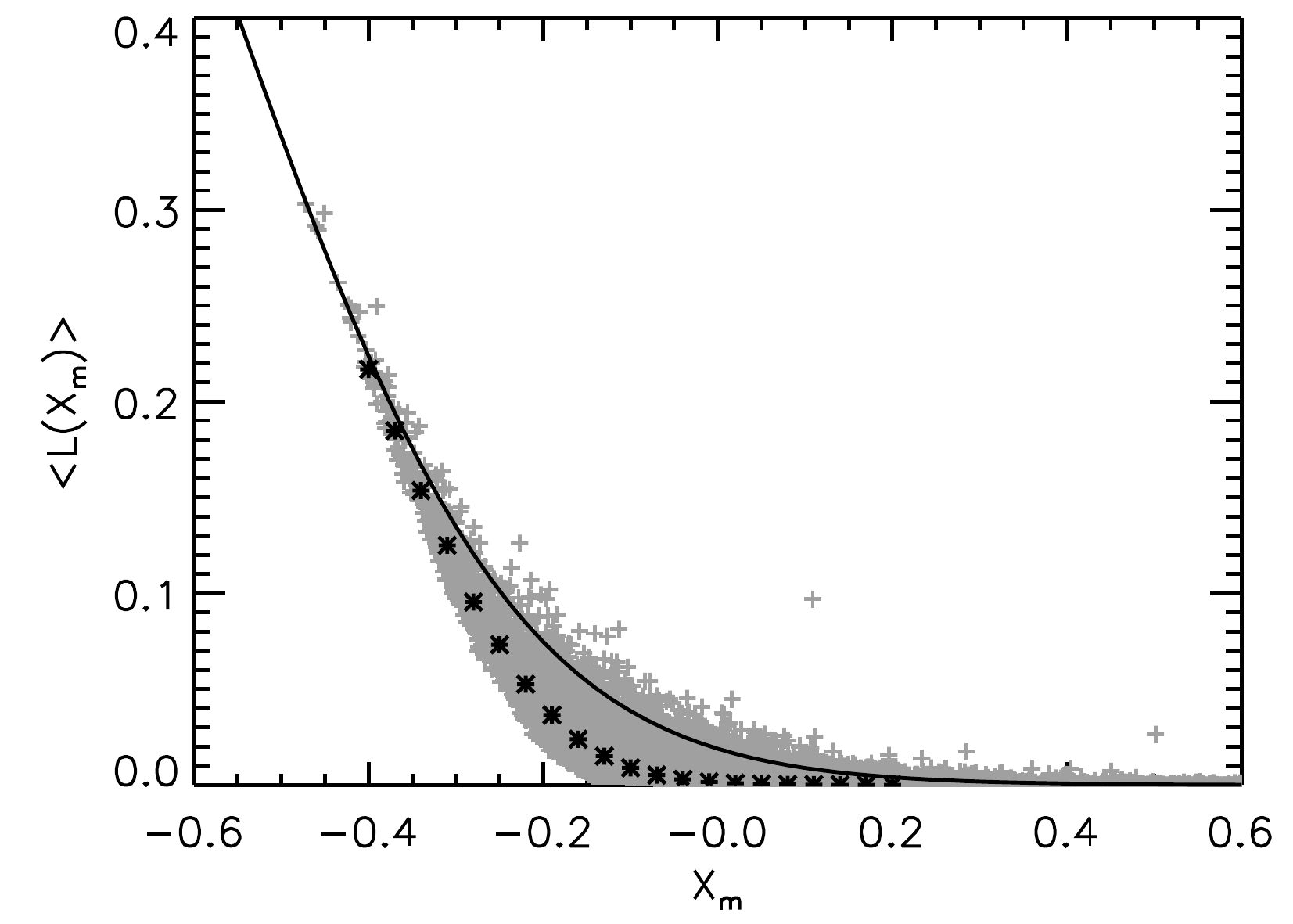}
\caption{
Dependence on market returns $X_\m$ for the GARCH process.
The left plot shows the dependence of default probabilities $\PD(X_\m)$ on market returns $X_\m$. 
The right plot shows the dependence of portfolio losses $\langle L(X_\m) \rangle$ on market returns $X_\m$.
The Monte Carlo results for the jump--diffusion process (grey symbols) are compared to the analytical results for the diffusion (solid line). The parameter $B$ is fitted to the default rate dependence in Fig.~\ref{fig:pddep_garch}.
}
\label{fig:xmdep_garch}
\end{center}
\end{figure*}

\begin{figure*}
\begin{center}
 \incfig{0.49\textwidth}{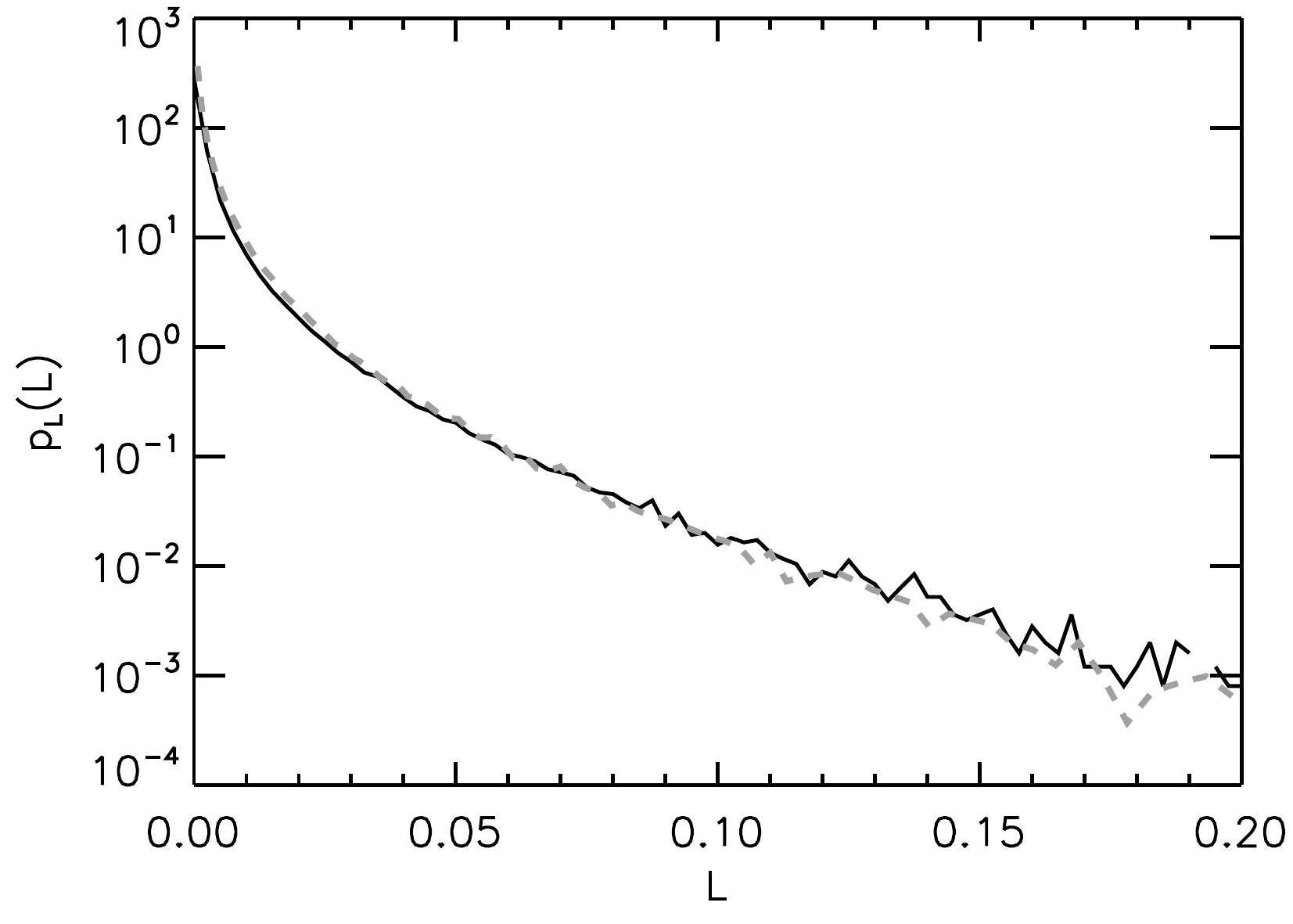}
\caption{
Probability density function $p_L(L)$ for the portfolio losses for the GARCH process.
The black solid line corresponds to the simulation results.
The grey dashed line shows the analytical results, where the pdf of default probabilities is transformed.
}
\label{fig:ploss_garch}
\end{center}
\end{figure*}

In Figure~\ref{fig:xmdep_garch} we demonstrate the dependence of default probabilities and portfolio losses on the market return. Here the behavior of the GARCH model is completely different from the diffusion case. For the analytical results of the diffusion case, described by Equations (\ref{eq:PDXm}) and (\ref{eq:plossrec1}), we use the parameter value $B$ which has been fitted to the $\PD$ dependence in Figure~\ref{fig:pddep_garch}.
Given the functional relation between portfolio loss $L$ and default probability $\PD$ we can transform the pdf of default probabilities into the pdf of portfolio losses,
\begin{equation}  \label{eq:pLpPD}
p_L(L)=\frac{1}{\left| L^\prime(\PD)  \right|} \, p_{\PD}(\PD)  \;.
\end{equation}
A comparison to the Monte Carlo result for the portfolio loss distribution is presented in Figure~\ref{fig:ploss_garch}. We observe a very good agreement, even for extremely large losses.
Finally, we present the results for Expected Loss, Value at Risk and Expected Tail Loss in Table~\ref{tab:garch}. 
We compare the Monte Carlo result for Value at Risk with
\begin{equation}
\mathrm{VaR}_{\alpha}=L(F_{\PD}^{-1}(\alpha)) \;,
\end{equation}
where $F_{\PD}^{-1}(\alpha)$ is the $\alpha$--quantile of default probabilities.
The Expected Tail Loss can be expressed as
\begin{equation}
{\rm ETL}_\alpha  
= \frac{1}{\alpha} \int\limits_{F_{\PD}^{-1}(\alpha)}^1 L(\PD)\, p_{\PD}(\PD) \, \rmd \PD  \;.
\end{equation}
The analytical values describe the simulation results very well. 

Our results demonstrate that the functional dependence of recovery rates on default probabilities does not depend on the underlying process. The analytical result for the diffusion process describes also other underlying processes in the Merton model. Hence we call Equation (\ref{eq:RPDrelation}) the {\it structural recovery rate}. Using this structural recovery rate we can describe the loss distribution if we have knowledge of the distribution of default probabilities.

\begin{table}[htdp]
\begin{center}\begin{tabular}{l c c} \hline
& \ simulation \ & \ analytical result  \\ \hline
EL \  &  $2.44 \cdot 10^{-3}$  &  $2.26 \cdot 10^{-3}$    \\
VaR$_{0.99}$ \ &  $3.28 \cdot 10^{-2}$  &     $3.16 \cdot 10^{-2}$   \\
ETL$_{0.99}$ \ &    $5.42 \cdot 10^{-2}$   &   $5.16 \cdot 10^{-2}$ \\ \hline
\end{tabular} \caption{Expected Loss, Value at Risk and Expected Tail Loss for the GARCH(1,1) process.}
\end{center}
\label{tab:garch}
\end{table}


\section{Applications of the structural recovery rate}  \label{sec6}


In a recent study \cite{Altman2005} find a strong negative correlation between default probabilities and recovery rates in empirical data.
%
The structural recovery rate provides such a negative relation between default and recovery rates.
It correctly describes the case of zero--coupon bonds for various underlying processes, 
but may even be applicable to a more general debt structure.
An indication for this is given by \cite{Chen2003} who show that the Merton model with jump--diffusion 
can be calibrated to provide the same yield spreads as reduced--form models.

Reduced--form models are also able to reproduce negative correlations between default and recovery rates, 
if the default model and the recovery model depend on a single common covariate, see \cite{Chava2008}.
Compared to the structural recovery rate, however, the reduced--form
recovery introduces more parameters and lacks a deeper motivation. 

In first passage models as described in \cite{Giesecke2004}, default may occur before maturity if the market value of the company falls below a default barrier. If defaults before maturity dominate in this model, recovery rates are independent from default rates.
The average recovery rate is then constant and does not resemble the negative relation mentioned above. 
However, if defaults occur mostly at maturity, the structural recovery rate is recovered.

\begin{figure*}
\begin{center}
 \incfig{0.49\textwidth}{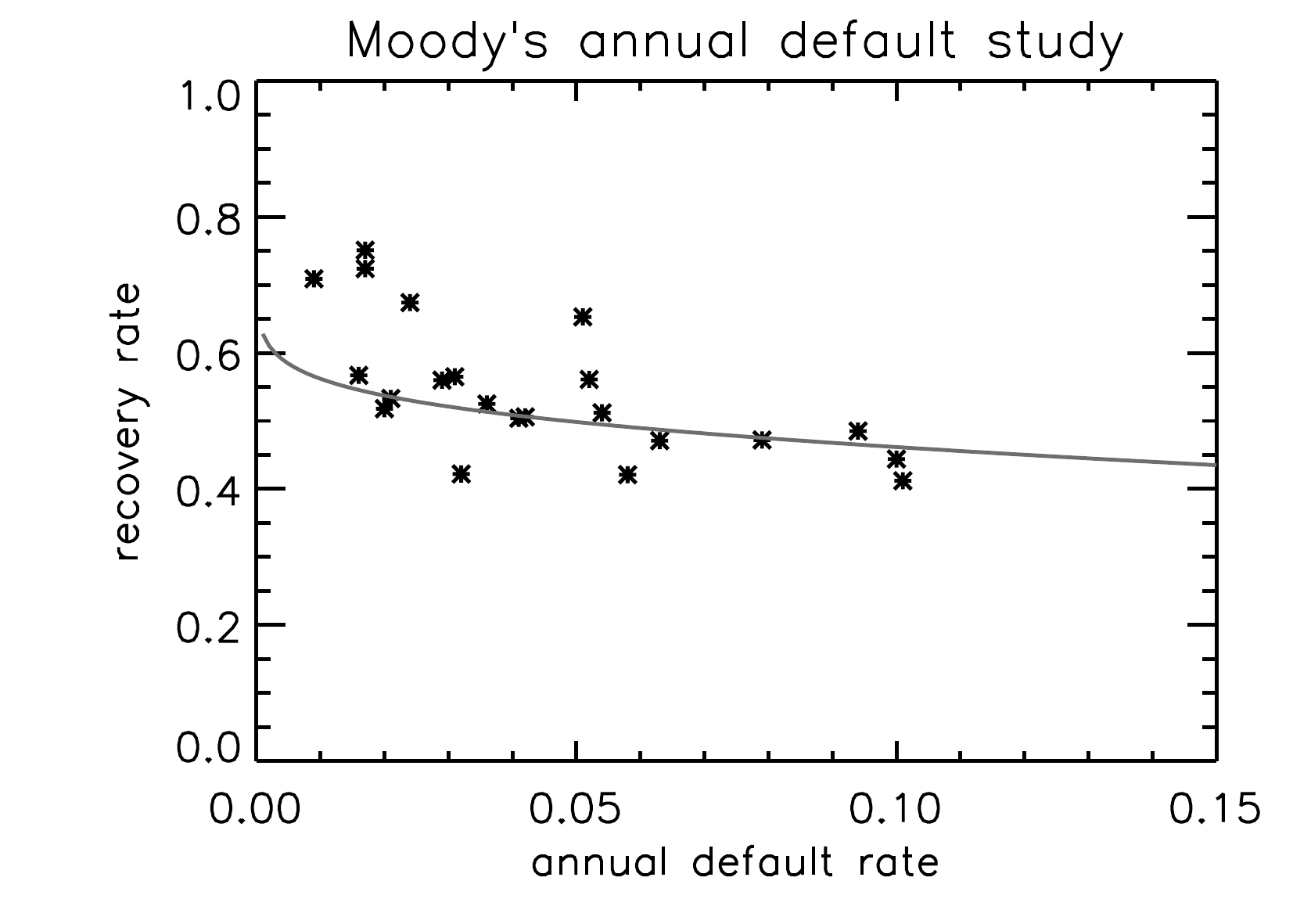}
 \incfig{0.49\textwidth}{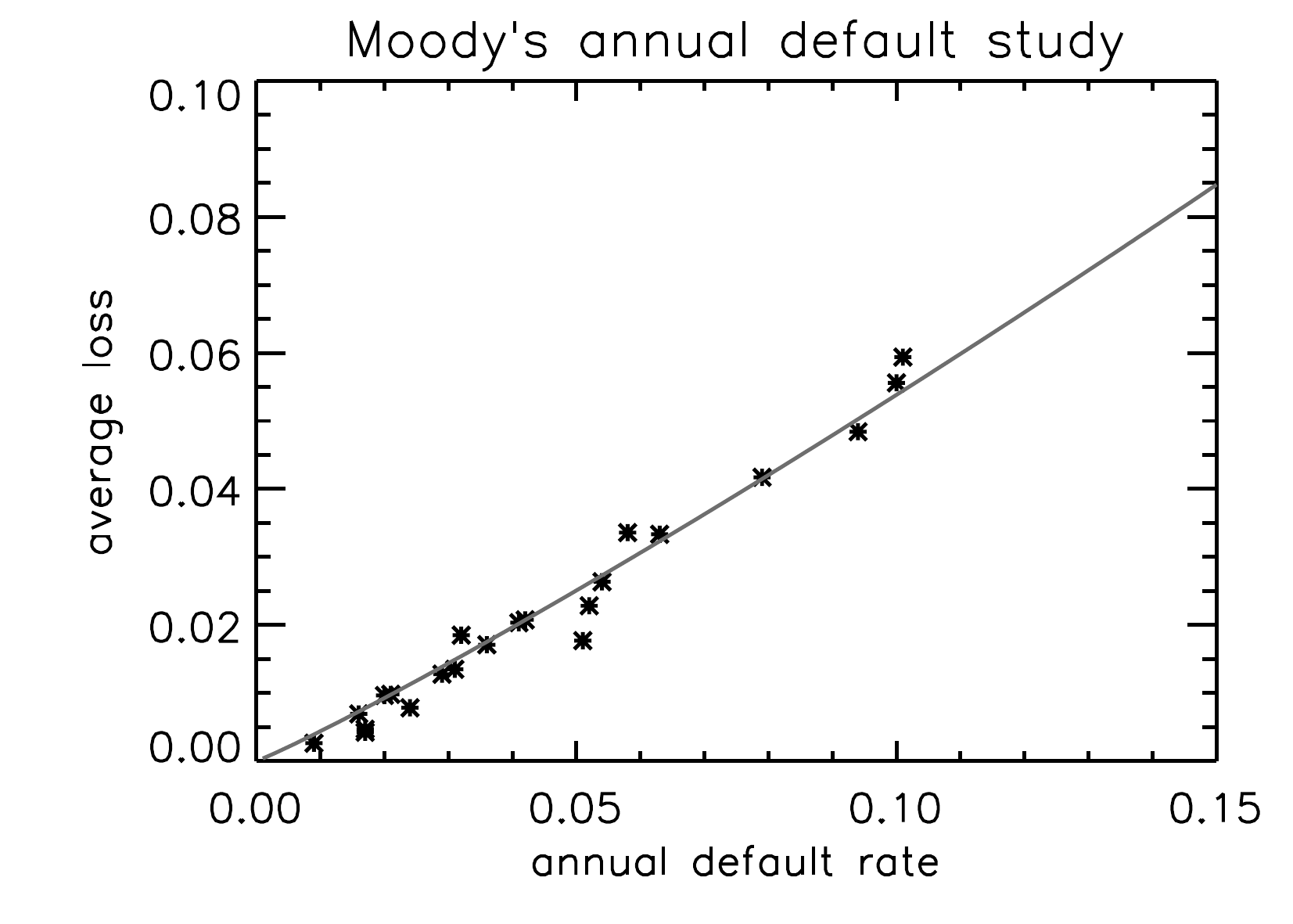}
\caption{
Empirical data from Moody's annual default study for the years 1987 to 2008.
The left plot shows the dependence of firm--wide ultimate recovery rates on the annual default rates. 
The right plot shows the dependence of the corresponding average losses  on the annual default rates.
The solid line shows the structural recovery rate with the parameter $B=2.28$ fitted to the data.
}
\label{fig:moodys}
\end{center}
\end{figure*}

Finally, we give a first empirical indication that the structural recovery rate is indeed applicable in a realistic setting.
Figure~\ref{fig:moodys} shows data from Moody's annual default study.
Although there are only very few data points, corresponding to the years 1987 to 2008, we can observe a negative relation between the firm--wide ultimate recovery rates and the annual default rates. We fit the structural recovery rate to these data points and find a reasonable agreement.


%
%




\section{Conclusions}  \label{sec7}

The interdependence of default and recovery rates has a crucial influence on large credit losses.
Yet default probabilities and recovery rates are often modelled independently in current credit risk models.
In this paper we revisited the Merton model for different underlying processes with correlations.
While the original Merton model was conceived for the diffusion process, we also consider a jump--diffusion and a GARCH process.
For the correlated diffusion we derived a functional dependence between default and recovery rates. This functional dependence is determined by a single parameter.
In Monte Carlo simulations we showed that it describes not only the diffusion case, but also the jump--diffusion and GARCH process.
Due to its independence on the underlying process, we call this functional relation the {\it structural recovery rate}.
It is straightforward to use this structural recovery rate in addition to a first passage model or any other model for default probabilities. 
We believe that this has great potential to improve current credit risk models.
In further studies we shall address two important questions: How well does the structural recovery rate describe empirical data on defaults and recoveries? 
And how does it compare with reduced--form models?

\section*{Acknowledgments}
We wish to thank Thomas Guhr and Sven {\AA}berg for helpful discussions.

\bibliographystyle{elsarticle-harv}
\bibliography{econophysics,creditrisk}

\begin{thebibliography}{20}
\expandafter\ifx\csname natexlab\endcsname\relax\def\natexlab#1{#1}\fi
\expandafter\ifx\csname url\endcsname\relax
  \def\url#1{\texttt{#1}}\fi
\expandafter\ifx\csname urlprefix\endcsname\relax\def\urlprefix{URL }\fi

\bibitem[{Altman et~al.(2005)Altman, Brady, Resti, and Sironi}]{Altman2005}
Altman, E.~I., Brady, B., Resti, A., Sironi, A., 2005. The link between default
  and recovery rates: Theory, empirical evidence, and implications. Journal of
  Business 78~(6), 2203--2228.

\bibitem[{Asvanunt and Staal(2009{\natexlab{a}})}]{AsvanuntStaal09a}
Asvanunt, A., Staal, A., April 2009{\natexlab{a}}. The Corporate Default
  Probability model in Barclays Capital POINT platform (POINT CDP). Portfolio
  Modeling, Barclays Capital.

\bibitem[{Asvanunt and Staal(2009{\natexlab{b}})}]{AsvanuntStaal09b}
Asvanunt, A., Staal, A., August 2009{\natexlab{b}}. The POINT Conditional
  Recovery Rate (CRR) Model. Portfolio Modeling, Barclays Capital.

\bibitem[{Black and Cox(1976)}]{BlackCox1976}
Black, F., Cox, J.~C., 1976. Valuing corporate securities: Some effects of bond
  indenture provisions. Journal of Finance 31, 351--367.

\bibitem[{Black and Scholes(1973)}]{black73}
Black, F., Scholes, M., 1973. The pricing of options and corporate liabilities.
  Journal of Political Economy 81~(3), 637.
\newline\urlprefix\url{http://www.journals.uchicago.edu/doi/abs/10.1086/260062}

\bibitem[{Bluhm et~al.(2002)Bluhm, Overbeck, and Wagner}]{bluhm02}
Bluhm, C., Overbeck, L., Wagner, C., 2002. An introduction to credit risk
  modeling. Chapman \& Hall/CRC.

\bibitem[{Bollerslev(1986)}]{bollerslev86}
Bollerslev, T., 1986. Generalized autoregressive conditional
  heteroskedasticity. J. Econometrics 31, 307--327.

\bibitem[{Chava et~al.(2008)Chava, Stefanescu, and Turnbull}]{Chava2008}
Chava, S., Stefanescu, C., Turnbull, S., 2008. Modeling the loss distribution,
  working paper.

\bibitem[{Chen and Panjer(2003)}]{Chen2003}
Chen, C.-J., Panjer, H., 2003. Unifying discrete structural models and
  reduced-form models in credit risk using a jump-diffusion process. Insurance:
  Mathematics and Economics 33~(2), 357--380.

\bibitem[{Duffie and Singleton(1999)}]{DuffieSingleton99}
Duffie, D., Singleton, K., 1999. Modeling the term structure of defaultable
  bonds. Review of Financial Studies 12, 687--720.

\bibitem[{Giesecke(2004)}]{Giesecke2004}
Giesecke, K., 2004. Credit Risk Modeling and Valuation: An Introduction, 2nd
  Edition. Credit Risk: Models and Management. Risk Books, Ch.~16, p. 487.

\bibitem[{Hull and White(2000)}]{HullWhite2000}
Hull, J.~C., White, A., 2000. Valuing credit default swaps {I}: No counterparty
  default risk. Journal of Derivatives 8~(1), 29--40.

\bibitem[{Jarrow et~al.(1997)Jarrow, Lando, and
  Turnbull}]{JarrowLandoTurnbull97}
Jarrow, R.~A., Lando, D., Turnbull, S.~M., 1997. A markov model for the term
  structure of credit risk spreads. Review of Financial Studies 10~(2),
  481--523.

\bibitem[{Jarrow and Turnbull(1995)}]{JarrowTurnbull95}
Jarrow, R.~A., Turnbull, S.~M., 1995. Pricing derivatives on financial
  securities subject to default risk. Journal of Finance 50, 53--86.

\bibitem[{Kiesel and Scherer(2007)}]{KieselScherer2007}
Kiesel, R., Scherer, M., 2007. Dynamic credit portfolio modelling in structural
  models with jumps, preprint available at
  \verb+http://www.defaultrisk.com/pp_model170.htm+.

\bibitem[{Merton(1974)}]{Merton74}
Merton, R.~C., 1974. On the pricing of corporate dept: The risk structure of
  interest rates. Journal of Finance 29, 449--470.

\bibitem[{Sch{\"a}fer and Guhr(2010)}]{SchaeferGuhr2010}
Sch{\"a}fer, R., Guhr, T., 2010. Local normalization: Uncovering correlations
  in non--stationary financial time series. Physica A 389~(18), 3856--3865.

\bibitem[{Sch{\"a}fer et~al.(2007)Sch{\"a}fer, Sj{\"o}lin, Sundin, Wolanski,
  and Guhr}]{schaefer07a}
Sch{\"a}fer, R., Sj{\"o}lin, M., Sundin, A., Wolanski, M., Guhr, T., 2007.
  Credit risk --- a structural model with jumps and correlations. Physica A
  383~(2), 533.

\bibitem[{Sch{\"o}nbucher(2003)}]{Schoenbucher2003}
Sch{\"o}nbucher, P.~J., 2003. Credit Derivatives Pricing Models. John Wiley \&
  Sons, New Jersey.

\bibitem[{Zhou(2001)}]{zhou01}
Zhou, C., 2001. The term structure of credit spreads with jump risk. Journal of
  Banking and Finance 25, 2015.

\end{thebibliography}

\end{document}